\documentclass[manuscript,11pt]{aastex}
\usepackage{epsfig}

\def\mathB{\textbf{\em B}}

\def\cm{\ifmmode {\rm cm}^{-1} \else cm$^{-1}$ \fi}
\def\s{\ifmmode {\rm s}^{-1} \else s$^{-1}$ \fi}
\def\cc{\ifmmode {\rm cm}^{-3} \else cm$^{-3}$ \fi}
\def\cs{\ifmmode {\rm cm}^{-2} \else cm$^{-2}$ \fi}
\def\g{\ifmmode \gamma \else $\gamma$\fi}
\def\G{\ifmmode \Gamma \else $\Gamma$\fi}
\def\Gs{\ifmmode \Gamma~ \else $\Gamma~$\fi}

\def\gc{\ifmmode \gamma_{\rm c} \else $\gamma_{\rm c}$ \fi}
\def\sw{Schwarzschild~}
\def\gsim{\mathrel{\raise.5ex\hbox{$>$}\mkern-14mu
             \lower0.6ex\hbox{$\sim$}}}
\def\lsim{\mathrel{\raise.3ex\hbox{$<$}\mkern-14mu
             \lower0.6ex\hbox{$\sim$}}}
\def\simless{\mathbin{\lower 3pt\hbox
     {$\rlap{\raise 5pt\hbox{$\char'074$}}\mathchar"7218$}}}   
\def\simmore{\mathbin{\lower 3pt\hbox
     {$\rlap{\raise 5pt\hbox{$\char'076$}}\mathchar"7218$}}}   
\def\Msun{M_\odot}                                
\def\4u{4U 1728--34}
\def\deg{^\circ}

\newcommand{\Alfven}{Alfv$\acute{\rm e}$n~}

\def\ark120{Ark~120}

\lefthead{et al.} \righthead{\4u}

\shorttitle{Soft Excess by GRMHD Hot Accretion}

\shortauthors{Fukumura et al.}

\begin{document}

\title{Soft X-Ray Excess from Shocked Accreting Plasma in Active Galactic Nuclei}

\date{\today}

\author{
\textsc{Keigo Fukumura}\altaffilmark{1,2,3},
\textsc{Douglas Hendry}\altaffilmark{1},
\textsc{Peter Clark}\altaffilmark{1},
\textsc{Francesco Tombesi}\altaffilmark{4,5},
\textsc{and}
\textsc{Masaaki Takahashi}\altaffilmark{6}
}

\altaffiltext{1}{Department of Physics and Astronomy, James Madison University, Harrisonburg, VA 22807}
\altaffiltext{2}{Email: fukumukx@jmu.edu}
\altaffiltext{3}{KITP Scholar at UC Santa Barbara }
\altaffiltext{4}{Astrophysics Science Division, NASA/Goddard Space Flight Center, Greenbelt, MD 20771}
\altaffiltext{5}{Department of Astronomy and CRESST, University of Maryland, College Park, MD20742} 
\altaffiltext{6}{Department of Physics and Astronomy, Aichi University of
Education, Kariya, Aichi 448-8542, Japan}

\begin{abstract}

\baselineskip=15pt

We propose a novel theoretical model to describe a physical identity of the soft X-ray excess, ubiquitously detected in many Seyfert galaxies, by considering a steady-state, axisymmetric plasma accretion within the innermost stable circular orbit (ISCO) around a black hole (BH) accretion disk. We extend our earlier theoretical investigations on general relativistic magnetohydrodynamic (GRMHD) accretion which has implied that the accreting plasma can develop into a standing shock for suitable physical conditions causing the downstream flow to be sufficiently hot due to shock compression. We numerically calculate to examine, for sets of fiducial plasma parameters, a physical nature of fast MHD shocks under strong gravity for different BH spins. We show that thermal seed photons from the standard accretion disk can be effectively Compton up-scattered by the energized sub-relativistic electrons in the hot downstream plasma to produce the soft excess feature in X-rays. As a case study, we construct a three-parameter Comptonization model 
of inclination angle $\theta_{\rm obs}$, disk photon temperature $kT_{\rm in}$ and downstream electron energy  $kT_e$ to calculate the predicted spectra in comparison with a 60 ks {\it XMM-Newton}/EPIC-pn spectrum of a typical radio-quiet Seyfert 1 AGN, Ark~120. Our $\chi^2$-analyses  demonstrate that the model is plausible in successfully describing data for both non-spinning and spinning BHs with the derived range of $61.3~{\rm keV} \lesssim kT_e \lesssim 144.3~{\rm keV}$, $21.6~{\rm eV} \lesssim kT_{\rm in} \lesssim 34.0~{\rm eV}$ and 
$17.5\degr \lesssim \theta_{\rm obs} \lesssim 42.6\degr$ indicating a compact Comptonizing region of $3-4$ gravitational radii that resembles the putative X-ray coronae.  



\end{abstract}



\baselineskip=15pt

\section{Introduction}

A broad-band synergistic study of active galactic nuclei (AGNs) in recent years has revealed a number of underlying spectroscopic components that is critical to understand a fundamental physical process around central engines of AGNs. Especially in a close proximity to a nucleus well within the sphere of influence  by a supermassive black hole (BH), state-of-the-art X-ray spectroscopies today have clearly demonstrated a complexity of AGN physics associated with inflows as well as outflows. Among others is a ``soft X-ray excess"  which is an excessive amount of spectral feature below $\sim 2$ keV above a baseline continuum  extrapolated from a hard X-ray range (typically from a $2-10$ keV power-law component of photon index $\Gamma \sim 2$), and it is known to be present among Seyfert galaxies particularly in the so called the  narrow-line Seyfert 1 (NLS1) AGNs --- a sub-class of the type 1 Seyfert galaxies of certain spectroscopic properties. Since its first extensive analysis as a part of the multi-wavelength campeign (EUV and soft/hard X-ray with {\it ROSAT}) of NLS1s about two decades ago \citep[e.g.][]{OsterbrockPogge85,Goodrich89,WalterFink93,Boller96,Leighly99a,Leighly99b}, only a small pieces of observational facts have been obtained even from more recent observations about the origin of soft excess; (1) Its presence is almost ubiquitous in NLS1 galaxies of both smaller and larger BH masses. (2)   Its equivalent blackbody temperature appears to be almost universally $\sim 0.1-0.2$ keV regardless of the objects. (3) Its output power occupies a good fraction of a total AGN luminosity \citep[][]{Boller96}.           
In the context of the standard accretion disk \citep[][]{SS73} its maximum effective temperature is well-defined to be only $kT \sim 10 (\dot{m}/M_8)^{1/4}$ eV where $\dot m \equiv \dot{M} / \dot{M}_{\rm Edd}$ is the AGN mass-accretion rate normalized by the Eddington mass-accretion rate for a BH mass $M$ with $M_8 \equiv M/(10^8 \Msun)$. Not only is it challenging to account for the observed nearly-constant  ``temperature" of the soft excess by the standard disk model \citep[e.g.][]{SS73}, the model, even with an extreme assumption of high accretion rate and low-mass BH, could barely bring the peak of disk spectrum only up to $kT \lesssim 0.1$ keV, which means that the standard disk emission should make virtually no significant contribution to the observed soft excess \citep[e.g.][]{Laor97} thus fails to explain its physical origin\footnote[1]{On the other hand, a slim disk configuration - another regime of accretion disks - has been also studied for NLS1s \citep[e.g.][]{Mineshige00}.}. While long sought, the nature of this spectral component is quite elusive so far. 

A number of plausible scenarios to explain the physics of the soft excess, among others,  includes  (1) 
%
a continuum component strongly absorbed by a series of  ionized absorbers in a relativistic outflow whose spectral curvature could then be interpreted as a falsified  ``excess" feature \citep[][]{SD06,SD08,GierlinskiDone04,Middleton07},  (2) ionized atomic processes from an inner part of the disk  illuminated via light bending \citep[e.g.][]{Miniutti04,Fabian04,Kara15} to produce a series of relativistically-blurred emission lines to mimic an apparently smooth ``excess" spectral shape \citep[e.g.][]{RossFabian05,Crummy06,Fabian09,Ponti10,Nardini11,DeMarco13} with a predicted correlation between the hard and soft X-rays \citep[e.g.][Boissay et al.~2015 in prep]{Vasudevan14} and (3) a Comptonization of the disk photons by some means such as corona ($kT=1$ keV) or upper layer of the disk \citep[e.g.][]{Petrucci04,Mehdipour11,Done12,ZW13,Noda13,DiGesu14} with an expectation of a correlation between UV and soft X-ray flux \citep[][]{Mehdipour11}. 
In particular, \citet{Petrucci13}, for example, has made a synergistic spectral analysis (from UV to hard X-ray) based on the multiwavelength campaign on a bright Seyfert galaxy, Mrk~509, focusing in part on the observed soft excess. Motivated by the implied correlation between UV and soft X-ray flux from {\it XMM-Newton} and INTEGRAL observations, the authors proposed a thermal Comptonization model to describe the physical origin of both soft excess and power-law components.


On the other hand, accretion physics has been extensively studied for decades particularly from theoretical aspects including semi-analytic investigations as well as global numerical simulations in an effort to further understand its physical nature and observational consequences. 
Many of these works on BH accretion in general have broadly revealed, among others, an  important generic feature of accretion; i.e. the formation of shocks 
%
%
as accreting plasma is subject to outward forces via a number of decelerating mechanisms \citep[e.g.][Abramowicz \& Prasanna (1990, MNRAS, 245, 720)]{} and develop a shock front at $r=r_{\rm sh}$ within the radius of the inner edge of a magnetized accretion disk\footnote[2]{\citet{Armitage01} has found an ISCO-like edge in their pseudo-Newtonian MHD accretion simulations. }, perhaps equivalent to a stable circular orbit (ISCO) for a pure HD Keplerian disk,  before crossing an event horizon at $r=r_H$. 
Previous studies include hydrodynamic shocks \citep[e.g.][]{NobutaHanawa94,Lu97,C90,F04} and magnetohydrodynamic (MHD) shocks (e.g., \citealt{Koide98,Koide00,Das07}; \citealt{T02}, hereafter T02; \citealt{T06}, hereafter T06; \citealt{F07}, hereafter F07; \citealt{FK07b}; \citealt{T10}).  
In particular, an extensive theoretical studies of various types of shocks have been conducted to date in an attempt to understand its  dynamical behavior; e.g. shock oscillation in the context of quasi-periodic oscillations (QPOs) and its spectroscopic signatures \citep[e.g.][]{CT95,Molteni96,Molteni99,Acharya02,Okuda04,Okuda07,NagakuraYamada08} that may be relevant for XRBs, for example. 
Independent GRMHD simulations of the tilted accretion disk clearly show that the compression of the plunging plasma in the inner region ($r \lesssim 10 r_g$) leads to the formation of standing shocks \citep[e.g.][]{Fragile07,FB08,Generozov14} depending on the characteristics of the disk geometry and the BH spin \citep[e.g.][]{Morales14}. The expected highly magnetized shocked region may perhaps correspond to the magnetically-arrested plasma  seen in the other large-scale simulations \citep[e.g.][]{TNM11}.

As a generic feature of accretion shocks unambiguously revealed in the earlier theoretical work by many authors, the downstream flow across the shock front is thus compressed and heated up efficiently to generate additional entropy  all the way down to the horizon unless a cooling process is sufficiently efficient \citep[e.g.][]{C95,CT95}.  While the detailed formalism and numerical methodology in these works are different, the presence of shocks in accretion is strongly favored in these calculations. As a result, the postshock region at small radii will provide an ideal site where the accelerated electrons could Compton up-scatter the thermal photons from an accretion disk. This process can produce a characteristic excess component in soft X-ray band below $\sim 1-2$ keV as a substantially modified disk blackbody radiation, and its  spectral shape depends on a number of variables related to MHD accretion processes. Our current work is thus motivated by this long-standing implication of shock formation in accretion.

\begin{figure}[ht]
\begin{center}$
\begin{array}{cc}
\includegraphics[trim=0in 0in 0in
0in,keepaspectratio=false,width=3.0in,angle=-0,clip=false]{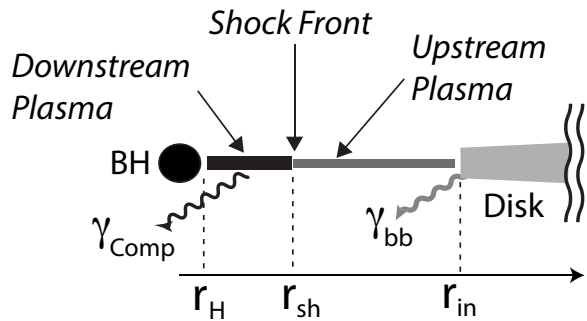}
\end{array}$
\end{center}
\caption{A schematic diagram illustrating nonthermal Comptonizing process of thermal disk photons of energy $kT_{\rm in}$ (labeled as $\gamma_{\rm bb}$) by  downstream energetic electrons of energy $kT_e$  to produce the Comptonized photons (labeled as $\gamma_{\rm Comp}$). See texts for the radii labelled. } \label{fig:f1}
\end{figure}

Utilizing the models of T02 and F07, we thus make a preliminary attempt in this paper to calculate the expected soft X-ray excess spectrum in the context of the well-explored GRMHD shocked accretion models in the past work (T02, F07). 
As depicted in Figure~\ref{fig:f1}, we  assume magnetized equatorial accretion with its inner edge of $r_{\rm in}$ (similar to the standard Shakura-Sunyaev disk \citealt{SS73}) in this model as a reservoir for incoming EUV photons. 
Plasma near $r=r_{\rm in}$ (i.e. ISCO-like boundary radius as discussed in \citealt{Armitage01})  eventually begins to plunge in and subsequently develops an adiabatic standing shock at a very small radius determined primarily by the plasma conditions such as energy, angular momentum and mass-accretion rate, for example. It is reminded, in the case of GRMHD accretion, that there exist a multiple magnetosonic points between $r_{\rm in}$ and $r_H$ (see T02 and T06 for more detailed discussion). 
The downstream region is therefore heated by adiabatic shock compression creating a centrally-concentrated,  compact hot region similar to the putative X-ray coronae \citep[e.g.][]{Fabian15,Wilkins15}.
The incoming blueshifted disk photons towards the downstream plasma is then Comptonized by hot electrons in the postshock flow producing the soft excess. 
Note, however, that the proposed Comptonization in our model is attributed exclusively to accelerated electrons in the rest-frame of downstream plasma accretion independent of the  bulk motion of the flow \citep[e.g., see,][for a fundamental difference from the standard bulk motion Comptonization model]{TMK96}.  
Our study focuses on explicitly constraining the defining parameters of the GRMHD accretion models by clearly identifying the physical origin of the observed soft excess feature as a shock-heating innermost downstream flow. 
Our ultimate goal is to systematically understand the physics of the  observed soft excess component in a coherent scenario by applying the model to a large sample of relevant Seyfet AGNs. As an example in this paper, we have analyzed a stereotypical radio-quiet Seyfert AGN, \ark120, as an application of the model and demonstrate its viability via $\chi^2$-statistics.     

We will briefly review the characteristics of the GRMHD accretion in \S 2 by presenting  fiducial shocked accretion solutions. Then, our methodology for calculating the Comptonized spectrum for a series of solutions in Kerr geometry is discussed in \S.3 with GR effects being fully implemented. In \S 4 we show our preliminary results for a 60-ks {\it XMM-Newton}/EPIC-pn spectrum of \ark120\ as a case study successfully constraining primary  variables in the model. We summarize and discuss the implications of the model in \S 5. 

\section{GRMHD Models with Shocked Accreting Plasma}

\subsection{Formalism}

We adopt the well-defined model for GRMHD shock formation in accreting plasma discussed in a series of papers (T02, T06, F07) in a formalism closely alined with other simulations, for example, by \citet{Pu15}. We consider stationary ($\partial_t=0$) and axisymmetric ($\partial_\phi=0$) ideal MHD accretion in Kerr geometry whose spacetime metric components $g_{\mu \nu}$ is described by the Boyer-Lindquist coordinates $(t, r,\theta,\phi)$
\begin{eqnarray}
ds^2 = \left(1-\frac{2Mr}{\Sigma}\right) dt^2 + \frac{4 M a r \sin^2 \theta}{\Sigma} dt d\phi - \frac{A \sin^2 \theta}{\Sigma} d\phi^2 - \frac{\Sigma}{\Delta} dr^2 - \Sigma d\theta^2 \ ,
\end{eqnarray}
with the conventional $(+---)$ metric signature where $M$ is BH mass and $a$ is its angular momentum per BH mass (i.e. spin parameter) with $\Delta \equiv r^2-2 M r+a^2$, $\Sigma \equiv r^2+a^2 \cos^2 \theta$, and $A \equiv (r^2 + a^2)^2-a^2 \Delta \sin^2 \theta$. The length scale in this paper is normalized to the gravitational radius $r_g$ where $r_g \equiv GM/c^2$ with $G$ and $c$ being the gravitational constant and speed of light, respectively.

In the context of ideal GRMHD, the property of accreting plasma is governed by (1) particle number
conservation law: $(n u^\alpha)_{;\alpha}$ where $n$ is the proper
particle number density and $u^\alpha$ is the plasma
four-velocity, (2) equation of motion:
$T^{\alpha\beta}_{~~;\beta}=0$ where $T^{\alpha \beta}$ is the
energy-momentum tensor for magnetized plasmas and (3) ideal MHD condition:
$u^\beta F_{\alpha \beta}=0$ where $F_{\alpha \beta}$ is the
electromagnetic field tensor. The poloidal plasma
four-velocity is given as $u_p^2 \equiv -u^\alpha u_\alpha$. The
energy-momentum tensor $T^{\alpha \beta}$ is given by
\begin{eqnarray}
T^{\alpha \beta} \equiv n \mu u^\alpha u^\beta -P g^{\alpha \beta}
+ \frac{1}{4 \pi} \left(F^{\alpha \gamma} F_{\
\gamma}^{\beta}+\frac{1}{4} g^{\alpha \beta} F^2 \right) \ ,
\label{eq:plasma-tensor}
\end{eqnarray}
where $F^2 \equiv F_{\mu \nu} F^{\mu \nu}$ and $\mu=(\rho+P)/n$ is
the relativistic enthalpy, $P$ is the thermal gas pressure, 
$\rho$ is the total energy density and $n$ the plasma number density. 
Note that we assume the polytropic relation as $P =  K \rho_o^{\Gamma_p}$ where $K$ is related to entropy of the plasma with the polytropic index $\Gamma_p$ and $\rho_o = n m_e$ is the rest-mass density with the particle rest-mass $m_e$.

Assuming a steady-state, axisymmetric plasma, one can describe a topology of magnetic field lines using the magnetic stream function $\Psi(r,\theta)$ which is constant along a given field line. The plasma is frozen-in and flows along the field lines with five constants of motion in this formalism; angular velocity of field lines $\Omega_F(\Psi)$, plasma flux to magnetic flux ratio $\eta(\Psi)$, total energy of accreting plasma $E(\Psi)$, total angular momentum of plasma $L(\Psi)$ and entropy $S(\Psi)$. The total energy and angular momentum of the adiabatic plasma are given by
\begin{eqnarray}
E &\equiv& \mu u_t -\frac{\Omega_F B_\phi}{4 \pi \eta}
\label{eq:E}
\ , \\
L &\equiv& -\mu u_\phi - \frac{B_\phi}{4 \pi \eta} \label{eq:L} \
,
\end{eqnarray}
where $B_\phi = (\Delta/\Sigma) F_{\theta r} \sin \theta$ is the toroidal component of the magnetic field seen by a distant observer. 
From the poloidal components of the equation of motion with the above five constants, one can derive the general relativistic Bernoulli equation (aka. poloidal/wind equation) as
\begin{eqnarray}
\mu^2 (1+u_p^2) = E^2 \left[ (\alpha-2M^2) f^2 - \delta \right]  \ ,
\end{eqnarray}
where $u_p^2 \equiv - (u_\alpha u^\alpha) = -(u_r u^r + u_\theta u^\theta)$ is the poloidal plasma velocity with $f, \delta$ and $\alpha$ being the functions of metric components and conserved quantities (see T06 and F07). Here, $M_A$ is the relativistic \Alfven Mach number\footnote[3]{The Mach number $M_A$ can decrease while plasma speeds up if magnetic field strength increases faster.} defined as
\begin{eqnarray}
M_A^2 \equiv \frac{4\pi \mu n u_p^2}{B_p^2} \ ,
\end{eqnarray}
where $B_p$ is the poloidal magnetic field in the distant observer's frame (e.g. T06; F07). 
Technically speaking, a field geometry should be self-consistently calculated by the force-balance equation in a direction parallel to the streamline described by the Grad-Shafranov (GS) equation in general relativistic regime.  However, we will adopt a simplistic approach (e.g. T02; F07) and specify a purely conical field line geometry such that the poloidal magnetic field is given by $|\mathB_p| \propto (\Delta \Sigma)^{-1/2}$ following the split-monopole approximation \citep[e.g.][; see also \S 5]{Michel73,Wald74,BZ77}. 
In this formalism, the field topology is thus parameterized to be conical.  
Plasma is assumed to be adiabatic of single temperature and we ignore its self-gravity and viscous nature. 

To describe the plasma kinematics from an intuitive perspective, we calculate and express the three-velocity components of plasma in two different locally-flat inertial frames \citep[e.g.][]{Manmoto00}; i.e. radial component $v^r_{\rm CRF}$ defined in a corotating reference frame (CRF) where a local observer is corotating with the plasma such that
\begin{eqnarray}
v^r_{\rm CRF} \equiv \left(\frac{-u_r u^r}{1-u_r u^r} \right)^{1/2}  \ ,
\end{eqnarray}
and toroidal component $v^\phi_{\rm LNRF}$ defined in a locally non-rotating reference frame (LNRF) where a zero-angular-momentum-observer (ZAMO) is corotating with a BH such that 
\begin{eqnarray}
v^\phi_{\rm LNRF} \equiv \frac{A}{r^2 \Delta^{1/2}} (\Omega-\omega)   \ ,
\end{eqnarray}
where $\Omega \equiv u^\phi/u^t$ is the angular velocity of plasma and  the frame-dragging $\omega \equiv -g_{t\phi}/g_{\phi\phi}$ has been subtracted off in LNRF so one sees the intrinsic plasma rotation locally.
Note that in these reference frames it is guaranteed to have $v_{\rm CRF}^r \rightarrow 1$ and $v_{\rm LNRF}^\phi \rightarrow 0$ as $r \rightarrow r_H$ by definition (see \S 2.2 for results).
To characterize the magnetized nature of the plasma, we examine the magnetization parameter $\sigma$ defined in LNRF as the ratio of the outward Poynting flux to the inward net mass-energy flux of the accreting plasma such that 
\begin{eqnarray}
\sigma \equiv \frac{B_\phi g_{\phi \phi} (\Omega_F-\omega)}{4 \pi \eta \mu u^t \rho_w^2}    \ ,
\end{eqnarray}
where $\rho_w^2 \equiv g^2_{t\phi}-g_{tt} g_{\phi \phi}$. Thus, the sign of $\sigma$ in general can change as $B_\phi$ may switch its direction due to the global field geometry (see T02 and F07 for details).

Assuming that accreting plasma is initially injected from a plasma source (most likely near the inner edge of a magnetized accretion disk; see Fig.~\ref{fig:f1})  with its toroidal velocity being predominant (i.e. $v_{\rm CRF}^r < v_{\rm LNRF}^\phi$) onto a BH, physical plasma accretion must be trans-magnetosonic before reaching the event horizon going through two magnetosonic points (i.e. slow and fast magnetosonic points) and the \Alfven point. On the course of accretion, furthermore, accreting plasma is subject to various ``obstacles" to slow down inwards; e.g. gas pressure, radiation pressure, magnetic force, and centrifugal barrier, for example. Via nonlinear processes, flow can develop into a shock front at some radius detrmined by a certain physical condition. Considering a proper jump condition across the shock front, one can determine a physically valid shock location $r_{\rm sh}$. 
Note that both upstream and downstream plasma must be trans-magnetosonic on its own; i.e. the former (latter) must pass through the {\it outer (inner)} magnetosonic points and the \Alfven point. 
To simplify the problem we set the surface of a shock front to be normal to the magnetic field lines as in the previous calculations (e.g., T02, F07). 
Among different types of shocks, we consider here adiabatic (i.e. Rankine-Hugoniot) perpendicular shock conditions where cooling processes are so inefficient at the shock front that no energy (and angular momentum) is dissipated away. 
The condition can be analytically simplified as  
\begin{eqnarray}
\frac{1}{\Delta \Sigma} \left(\frac{M_1}{4\pi \eta} \right)^2 + \frac{\mu_1 -1}{1+N} \frac{\mu_1}{M_1^2} + \frac{\left(E f_1 \right)^2}{2} &=& \frac{1}{\Delta \Sigma} \left(\frac{M_2}{4\pi \eta} \right)^2 + \frac{\mu_2 -1}{1+N} \frac{\mu_2}{M_2^2}  + \frac{\left(E f_2 \right)^2}{2} \ ,
\end{eqnarray}
where the roots ($r=r_{\rm sh}$) to this equation are ``shock locations" and the subscripts ``1" and ``2"  respectively denote upstream and downstream quantities. We have used $1+N \equiv \Gamma_p/(\Gamma_p-1)$. 
We will numerically  calculate this radius across which  particle number, energy, angular momentum, and magnetic flux are all simultaneously conserved (while density, temperature and velocity are discontinuous). 
Note that the enthalpy $\mu$ remains conserved due to adiabatic assumption but increases across the shock because of entropy generation. 
The local shock compression in the Newtonian view is then given by the velocity ratio as
\begin{eqnarray}
\frac{n_2}{n_1} \equiv \frac{u^r_1}{u^r_2} = \frac{\mu_2 M_{A,1}^2}{\mu_1 M_{A,2}^2} \ .
\end{eqnarray}
As in the past work (e.g. T02; F07) we  simply treat the shock front as a mathematical discontinuity rather than considering its actual finite internal structure \citep[see, e.g.,][for considering particle transport at a diffusive shock front]{LeBecker05,Becker11}. 
It is reminded that the postshock downstream flow is heated across the shock front in the absence of efficient cooling mechanisms and  remain hot under adiabatic flow assumption in this work. 

Most importantly in the present work, we also introduce  normalized electron thermal energy $\Theta_e$  to assess energetics of accelerated electrons in the heated downstream plasma due to shock compression as a function of radius  as 
\begin{eqnarray}
\Theta_e(r) \equiv \frac{k T_e}{m_e c^2} = K \rho_o^{\Gamma_p-1} = \frac{1}{1+N} \left(\frac{\mu}{m_e c^2} -1 \right) \ ,
\end{eqnarray}
where $k$ is the Boltzman constant, $m_e c^2 = 511$ keV is the electron's rest-mass energy and $T_e$ is electron's equivalent thermal temperature in a single-fluid approximation. Hence, the temperature is uniquely determined by plasma density and entropy which is closely related to plasma pressure as well from the polytropic assumption. 
Unlike the hot downstream flow, the upstream (preshock) plasma is assumed to have negligible thermal energy (i.e. $K \sim 0$) compared to its rest-mass energy  in the cold flow limit (i.e. $\mu_1 \sim m_e c^2$ thus $\Theta_e \sim 0$ in the upstream region). In this limit, the slow magnetosonic point for the upstream plasma vanishes leaving only the \Alfven point and a fast magnetosonic point.  

\begin{figure}[ht]
\begin{center}$
\begin{array}{ccc}
\includegraphics[trim=0in 0in 0in
0in,keepaspectratio=false,width=2.0in,angle=-0,clip=false]{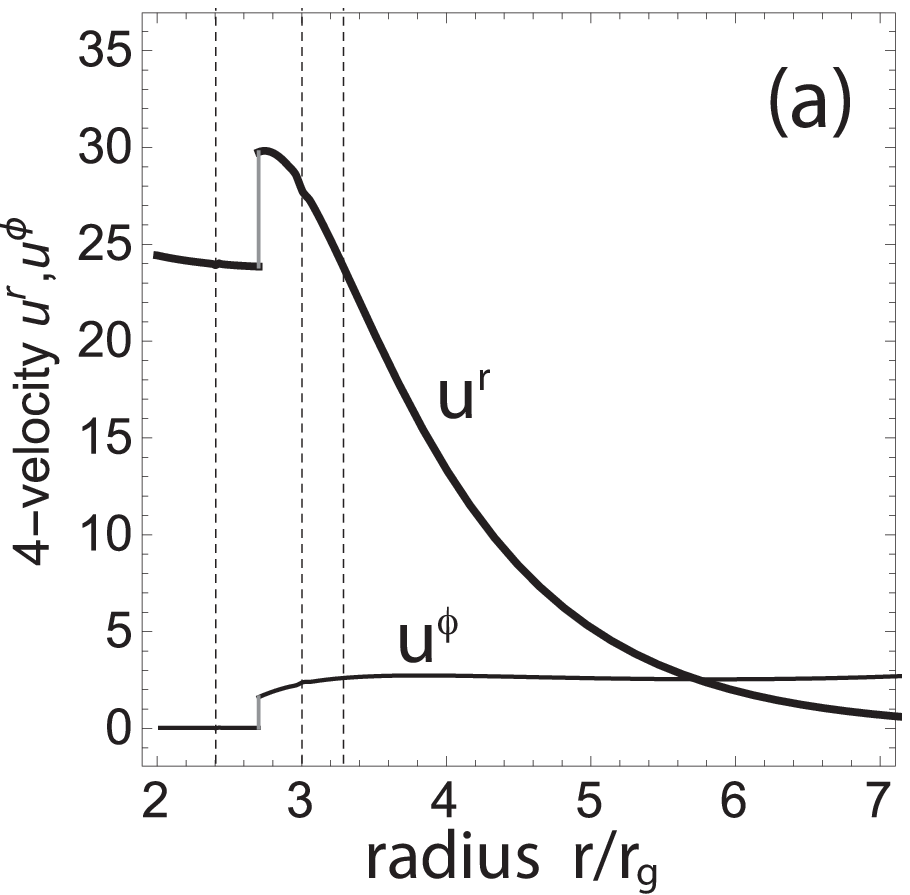}  &
\includegraphics[trim=0in 0in 0in
0in,keepaspectratio=false,width=2.0in,angle=-0,clip=false]{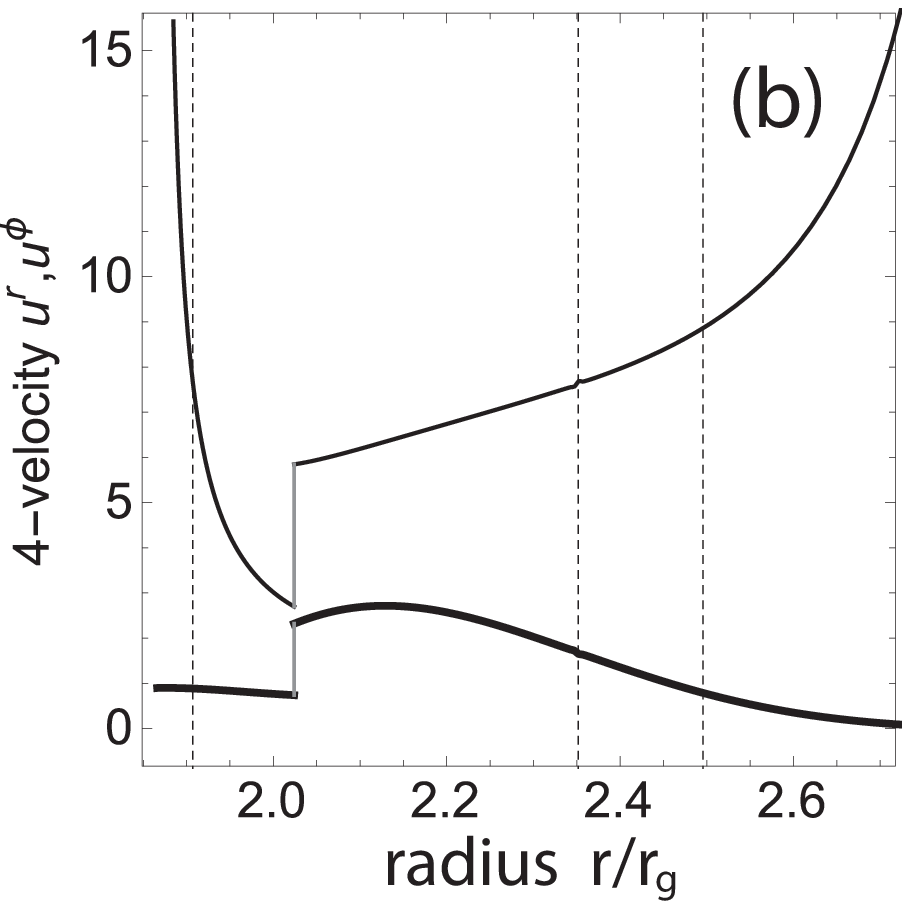} &
\includegraphics[trim=0in 0in 0in
0in,keepaspectratio=false,width=2.1in,angle=-0,clip=false]{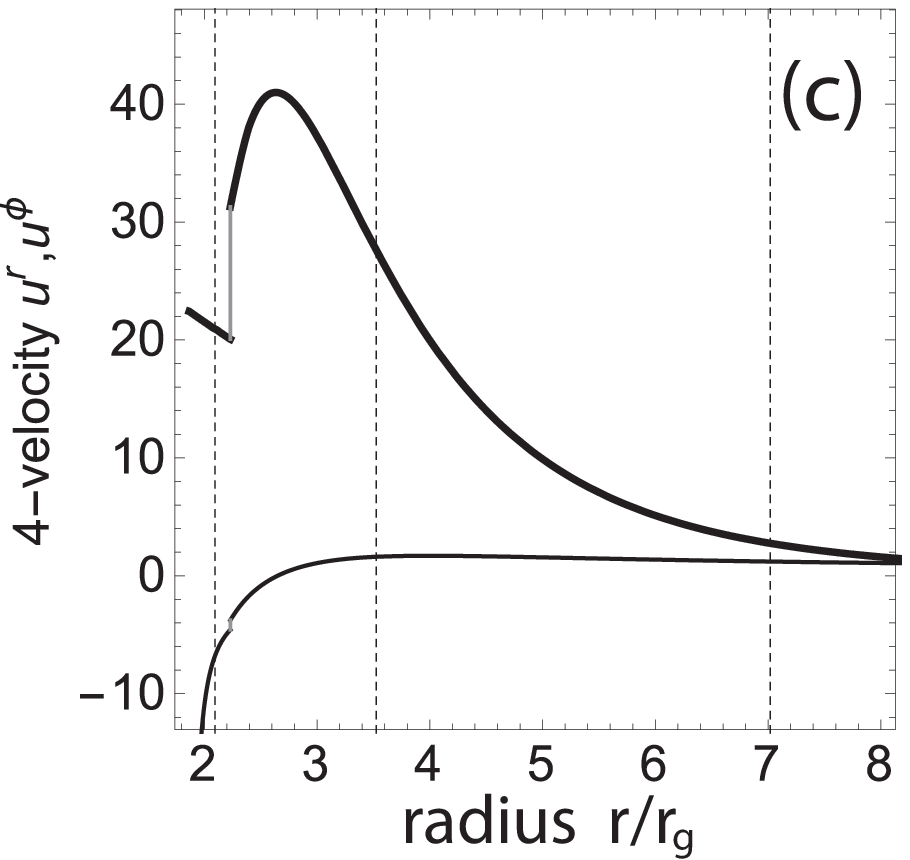}  
\end{array}$
\end{center}
\caption{Radial profiles of four-velocity components $u^r$ (thick solid) and $u^{\phi}$ (solid)  for a fiducial accreting MHD plasma with (a) $a/M=0$ (\sw BH), (b) $0.5$ (prograde) and (c) $-0.5$ (retrograde). Vertical dotted lines denote the outer \Alfven radius (rightmost), outer fast-magnetosonic radius (middle) and the inner \Alfven radius (leftmost) as plasma accretes and develops a fast MHD shock as shown in vertical jump. See Table~\ref{tab:tab1} for detailed model parameters. } \label{fig:f2}
\end{figure}

\begin{deluxetable}{l c ccc}
\tablecaption{Characteristics of Fiducial GRMHD Plasma Accretion \label{tab:tab1}}
\tabletypesize{\small}
\tablecolumns{10}
\tablewidth{0pt}
\tablehead{
	\colhead{Conserved Parameter} & \colhead{Description} &
	\multicolumn{3}{c}{BH Spin $a/M$} \\
	\cline{3-5} \\
	\colhead{} & \colhead{} &
	\colhead{-0.5} & \colhead{0} & \colhead{0.5} 
}
\startdata
$E$ & Energy & 6.1 & 6.1 & 6.1  \\ 
$L/E$ & Specific angular momentum & 2.1 & 2.3 & 3.96  \\  
$\Omega_F$ & Angular velocity of field line & 0.02725 & 0.08334 & 0.2333  \\
$4\pi \eta$ & Scaled accretion energy  & 0.0082 & 0.006 & 0.005 \\  \hline
$r^{\rm out}_A/r_g$ & Outer \Alfven radius & 7.01 & 3.28 & 2.82 \\
$r^{\rm out}_F/r_g$ & Outer fast radius & 3.52 & 3.01 & 2.49 \\
$r_{\rm sh}/r_g$ & Shock location & 2.23 & 2.70 & 1.99 \\
$r^{\rm in}_F/r_g$ & Inner fast radius & 2.09 & 2.41 & 1.90 \\
$r_H/r_g$ & Event horizon & 1.86  & 2.0 & 1.86 \\
$\Theta_e(r=r_{\rm sh})$ & Electron energy  & 0.358 & 0.199 & 0.285 \\
\enddata \\
\vspace{0.1cm}
Note: Superscripts ``in" and ``out" respectively denote those radii for the ``downstream" and ``upstream" plasma. 
\vspace*{0.2cm}
\end{deluxetable}

\clearpage

\subsection{Numerical Solutions for Shocked Plasma Accretion}

Following the past work (T02, F07), we calculate a global property of physically valid plasma accretion for a given set of conservative quantities described in \S 2.1. Our calculations throughout this paper are restricted to the equatorial flows for simplicity (i.e. $\theta=\pi/2$ and $u^\theta=B^\theta=0$) and  we set the polytropic index $\Gamma_p=4/3$ in the presence of the conical magnetic field. To exploit the parameter space as systematically as possible, a fiducial value of the plasma energy is chosen as $E=6.1$ in all cases discussed here (see F07 for its relevance) along with the other conserved quantities and radii as listed in Table~\ref{tab:tab1}.
Note that these characteristic radii are not free-parameters but determined by the shock conditions. 
We then vary the rest of the primary parameters, $L/E$ and $\Omega_F$ for a given BH spin $a$ and $\eta$ in order to find the valid solutions (see T02 and F07 for a detailed numerical methodology).

\begin{figure}[ht]
\begin{center}$
\begin{array}{ccc}
\includegraphics[trim=0in 0in 0in
0in,keepaspectratio=false,width=2.0in,angle=-0,clip=false]{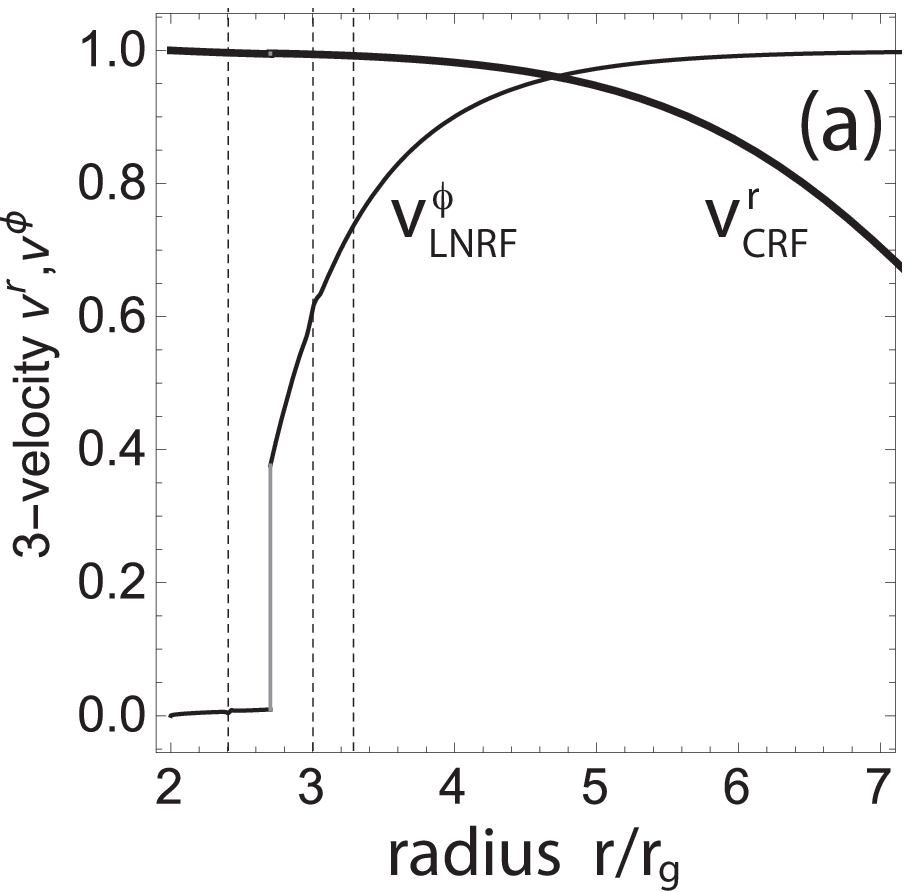} &
\includegraphics[trim=0in 0in 0in
0in,keepaspectratio=false,width=2.0in,angle=-0,clip=false]{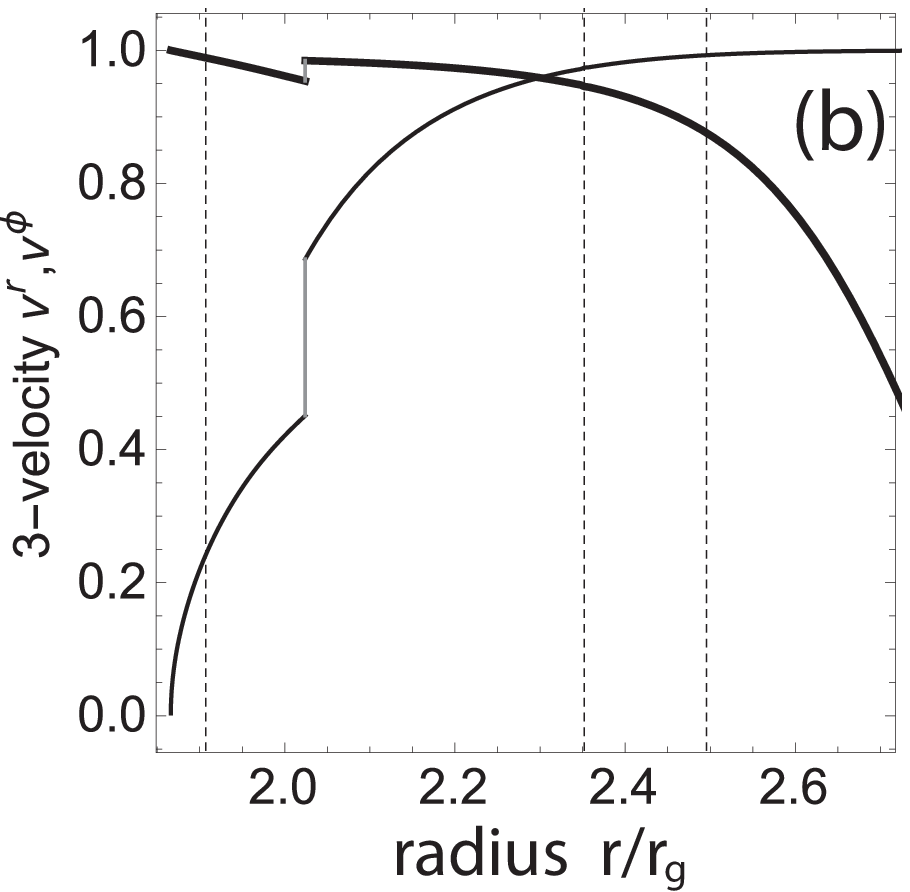} &
\includegraphics[trim=0in 0in 0in
0in,keepaspectratio=false,width=2.1in,angle=-0,clip=false]{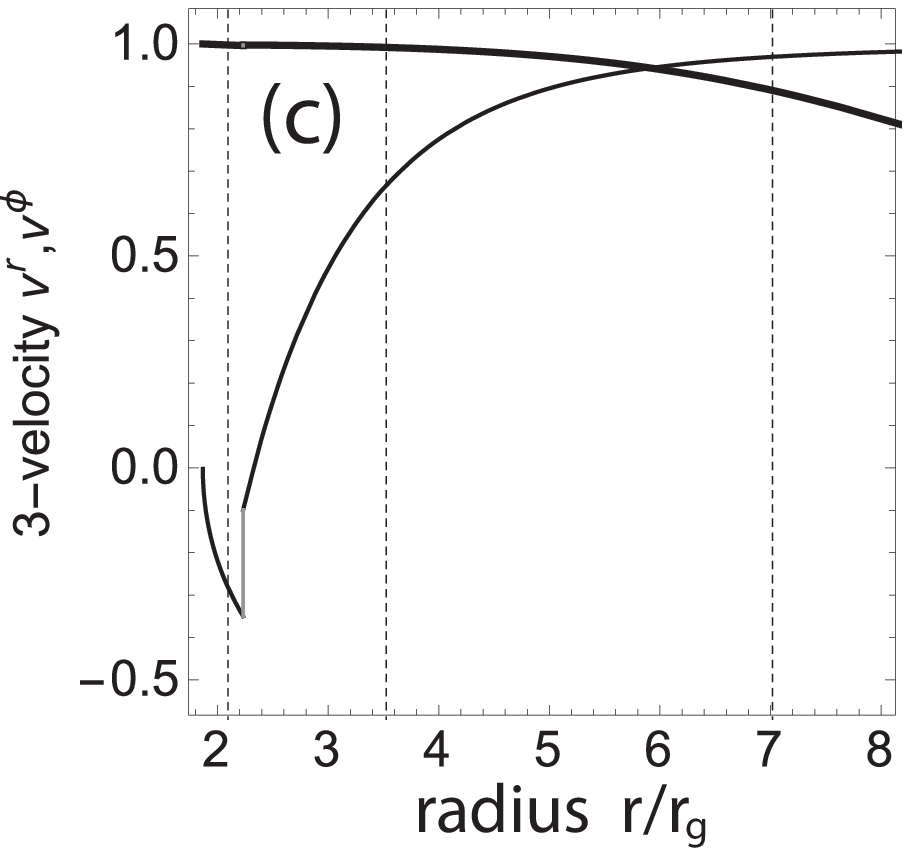} 
\end{array}$
\end{center}
\caption{Same as Figure~\ref{fig:f2} but for physical three-velocity components $v^r_{\rm CRF}$ (thick) and $v^{\phi}_{\rm LNRF}$ (solid)  for  accreting MHD plasma with (a) $a/M=0$ (\sw BH), (b) $0.5$ (prograde) and (c) $-0.5$ (retrograde).  } \label{fig:f3}
\end{figure}

\begin{figure}[ht]
\begin{center}$
\begin{array}{ccc}
\includegraphics[trim=0in 0in 0in
0in,keepaspectratio=false,width=2.0in,angle=-0,clip=false]{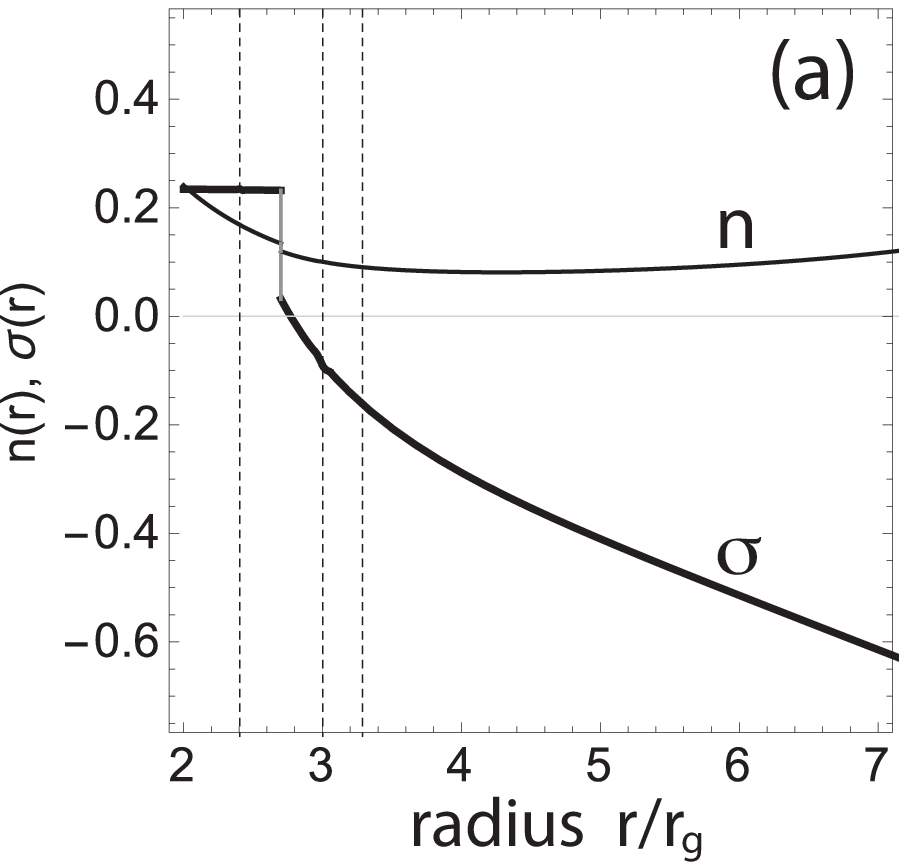} &
\includegraphics[trim=0in 0in 0in
0in,keepaspectratio=false,width=2.0in,angle=-0,clip=false]{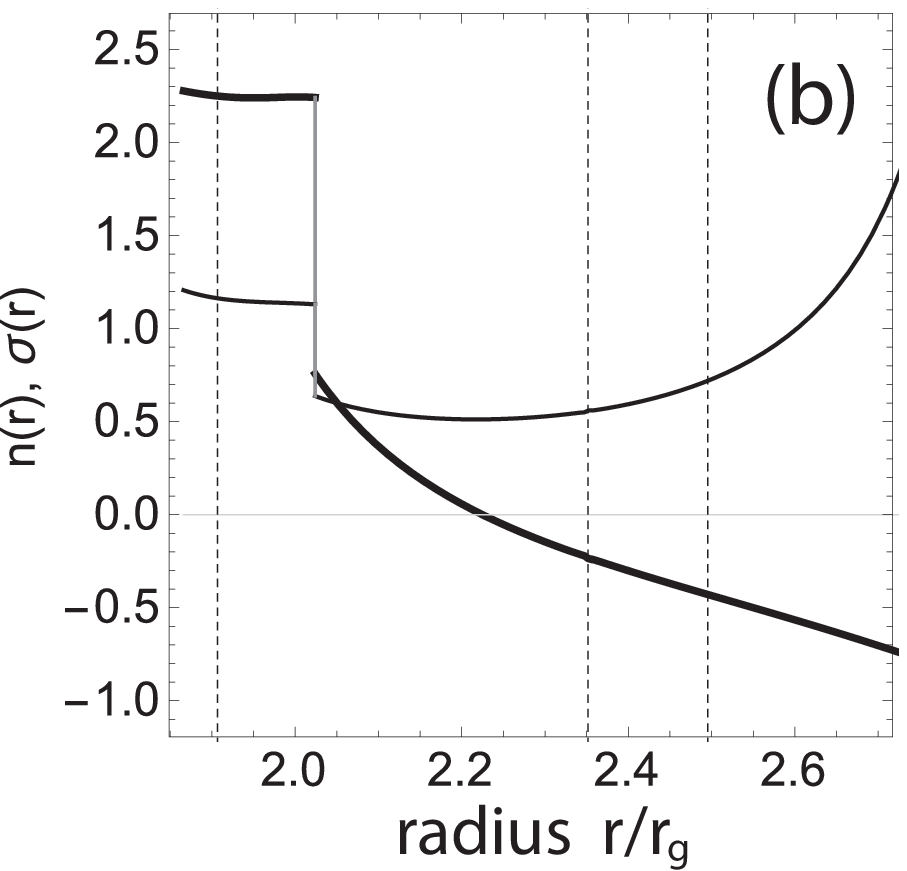} &
\includegraphics[trim=0in 0in 0in
0in,keepaspectratio=false,width=2.0in,angle=-0,clip=false]{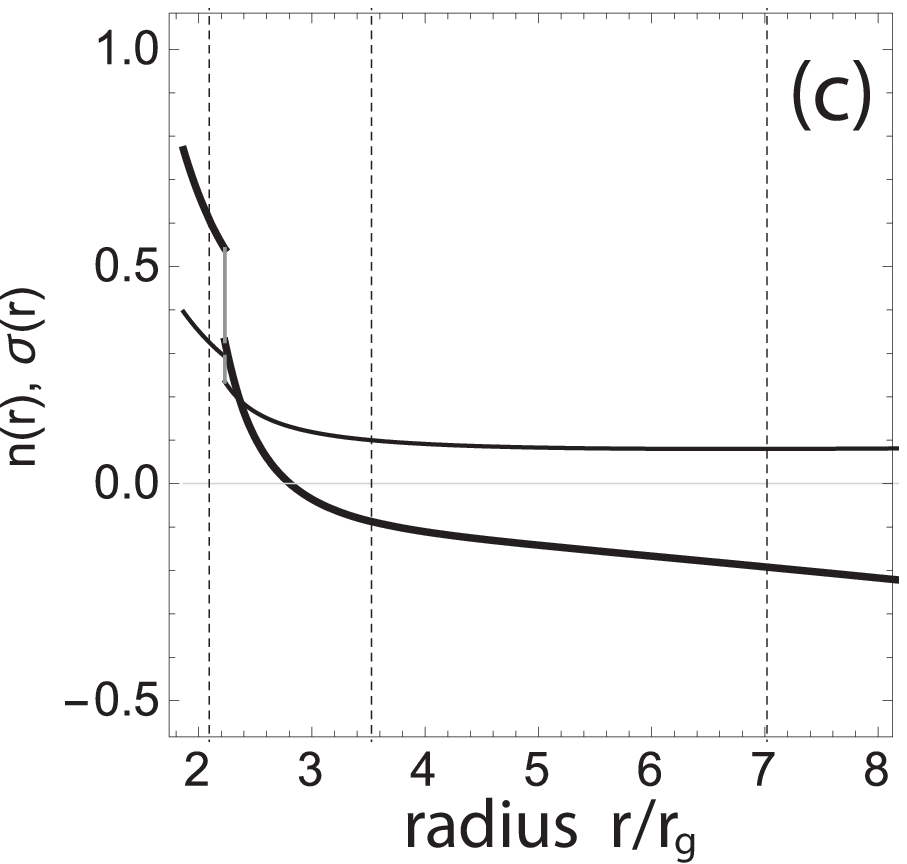} 
\end{array}$
\end{center}
\caption{Same as Figure~\ref{fig:f2} but for plasma magnetization $\sigma$ (thick) and number density $n$ (solid)  for  accreting MHD plasma with (a) $a/M=0$ (\sw BH), (b) $0.5$ (prograde) and (c) $-0.5$ (retrograde).  } \label{fig:f4}
\end{figure}

As a representative solution for a given BH spin, normalized radial profiles of major characteristics of each shocked accretion is shown in Figures~\ref{fig:f2}-\ref{fig:f4} for (a) $a/M=0$, (b) $0.5$ and (c) $-0.5$ along with its corresponding outer/inner fast-magnetosonic radii and the \Alfven radius (vertical dotted lines). A standing shock is denoted as a solid vertical line (gray) connecting upstream and downstream flows. Figure~\ref{fig:f2} shows the computed radial $u^r$ and azimuthal $u^\phi$  components of the four-velocity in the Boyer-Lindquist coordinates. The plasma begins to plunge in  near the ISCO along a magnetic field line radially accelerating by passing through the \Alfven point and the outer fast point. The upstream flow then forms a shock at $r=r_{\rm sh}$ developing a hot downstream region followed by passing through the inner fast point before entering the event horizon in each BH spin case (a)-(c). Note that the upstream plasma starts slowing down due to a number of outward forces (e.g. Lorentz force and centrifugal barrier) just before forming a shock. In the case of a \sw BH in (a), the plasma motion becomes more and more radial towards the horizon (i.e. $u^r > u^\phi$). For a prograde BH in (b), the plasma starts to plunge in from the ISCO at smaller radius compared to (a) because the ISCO radius shifts more inward. Therefore, the plunging plasma acquires a larger toroidal velocity $u^\phi$ due to a faster  Keplerian motion of the disk in the beginning.  
At small radius, the frame-dragging effect is prominent forcing the downstream flow to be corotating with the BH and $u^\phi$ dominates over $u^r$. On the other hand, around a retrograde BH in (c), the rotational sense of the upstream plasma ($u^\phi>0$ at large radii) is eventually switched the other way around ($u^\phi<0$ near the horizon) due to the frame-dragging, as expected. In other words, a distant observer in a flat spacetime sees the accreting plasma momentarily turn around at some point (i.e. $u^\phi=0$) between the outer fast point and the shock location  changing the direction of its toroidal motion.  

A corresponding three-velocity of the plasma is calculated in a local reference frame of flat spacetime as shown in Figure~\ref{fig:f3}. In all the cases the plasma is seen in CRF to radially approach the speed of light ($v^r_{\rm CRF} \rightarrow 1$) in the course of accretion while toroidal motion seen in LNRF eventually approach  the spacetime rotational speed ($\Omega \rightarrow \omega$ thus $v^\phi_{\rm LNRF} \rightarrow 0$) as expected. Note in (c) that the same  ``turn-around" behavior of the plasma toroidal motion is clearly seen in LNRF between the outer fast point and the shock location (i.e. $v^\phi_{\rm LNRF}>0 \rightarrow v^\phi_{\rm LNRF}<0$) due to the frame-dragging and in fact the plasma ``overshoots" the ZAMO in LNRF until the shock occurs (thus $v^\phi_{\rm LNRF}$ continues to increase in the same sense as the BH rotation). In the downstream flow, the shocked plasma stops ``overshooting" the ZAMO  eventually converging to the frame-dragging at the horizon as expected. It is counterintuitive to see in (c) that the plasma seemingly appears to ``speed up" in toroidal direction across the shock (i.e. $|v^\phi_{\rm LNRF}(r)|$ increases across the shock) in LNRF although in radial direction the plasma indeed slows down (i.e. $v^r_{\rm CRF}$ decreases across the shock while hard to see in the plot) in CRF. This is explained as follows; we see $\Omega>0$ and $\omega<0$ at large radii in the upstream flow (i.e. $v^\phi_{\rm LNRF}>0$).  Plasma is inevitably forced to slow down as it accretes due to frame-dragging and at some point it appears to come to a stop momentarily (i.e. $\Omega=0$) with respect to the observer (i.e. still $v^\phi_{\rm LNRF} >0$). The plasma then turns around in the same sense as the BH rotation (i.e. still $\omega < \Omega < 0$ thus $v^\phi_{\rm LNRF}>0$ again) and later converges to the frame-dragging (i.e. $\Omega=\omega <0$ thus $v^\phi_{\rm LNRF}=0$). Because of the initial angular momentum, the plasma subsequently ``overshoots" in toroidal direction allowing for $\Omega < \omega < 0$ thus $v^\phi_{\rm LNRF}<0$. Past this phase, the plasma starts to converge to the BH rotation at small radii and eventually acquire $\Omega=\omega<0$ thus $v^\phi_{\rm LNRF}=0$.

Besides plasma kinematics, magnetization $\sigma(r)$ and plasma number density $n(r)$ are shown in Figure~\ref{fig:f4}. Across the shock the upstream flow becomes compressed in all cases by definition causing the downstream flow to be heated where particles (primarily electrons) can be efficiently accelerated to later participate in Comptonization (see \S 3).   
Magnetization parameter is initially negative at large radii because the Poynting flux in the upstream flow is directed radially outward while accretion energy flux always points inward. The Poynting flux is then shifted inward because $B_\phi$ switches its direction due to the curvature of the field line (see T02). 
In terms of energy budget of the plasma flow across fast MHD shocks, a fraction of upstream accretion energy is redistributed to both magnetic field and thermal energies and hence the magnetization always increases (via increasing $B_\phi$) and entropy generation (via $K$) due to shock compression.  The downstream region is therefore always more magnetized (i.e. $|\sigma_2| > |\sigma_1|$).

\begin{figure}[ht]
\begin{center}$
\begin{array}{ccc}
\includegraphics[trim=0in 0in 0in
0in,keepaspectratio=false,width=3.0in,angle=-0,clip=false]{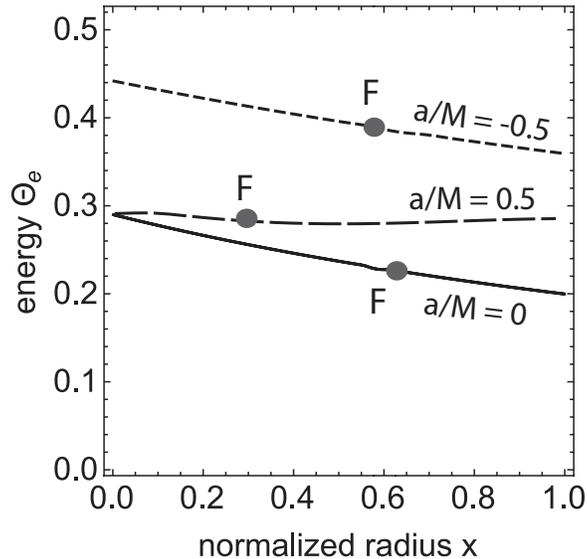} 
\end{array}$
\end{center}
\caption{Same as Figure~\ref{fig:f2} but downstream electron energy $\Theta_e \equiv kT_e/(m_e c^2)$  for  accreting plasma with $a/M=0$ (solid), $0.5$ (dotted) and $-0.5$ (dashed) where $x \equiv (r-r_{H})/(r_{\rm sh}-r_H)$ is  a normalized radius  between the horizon $r_H$ and the shock location $r_{\rm sh}$ such that $x = 0$ at the horizon $r_H$ while  $x=1$ at the shock location $r_{\rm sh}$ as defined in the text. A letter ``F" denote the corresponding inner fast-point in each case. } \label{fig:f5}
\end{figure}

Finally, we show  in Figure~\ref{fig:f5} the downstream thermal energy $\Theta_e(r)$ of the trans-magnetosonic plasma as a function of normalized radius $x \equiv (r-r_H)/(r_{\rm sh}-r_H)$ where $r_{\rm sh}$ is the shock location. As seen, the shock heating can raise the plasma thermal energy up to $\lesssim 40\%$ of the rest-mass energy of the plasma. 
It is found that $\Theta_e(r)$ only slowly varies with radius in almost all cases because (i) it is closely related to the density $n(r)$ which is almost constant in the postshock plasma (except for the retrograde case in (c)), (ii) the shock heating is not dissipated due to adiabatic assumption but advected and (iii) small radial size of the downstream flow. We will discuss later in \S 3 that the value of $\Theta_e(r)$ plays a fundamental role in determining the degree of Comptonization of incoming thermal disk photons. 

\section{Bulk Comptonization in the Hot Plasma}

\subsection{Comptonizing Process in the Downstream Plasma}

Across a shock front 
%
electrons in accreting flow are efficiently compressed and heated up to sub-relativistic regime (i.e. $kT_e/m_e c^2 < 1$) in the postshock flow possibly via the first-order Fermi mechanism \citep[e.g.][]{Fermi49,Baring97,GieselerJones00,LeBecker05} in the presence of the randomly distributed, turbulent magnetic fields. 
Assuming that the energy due to shock heating $\Theta_e(r)$ is fully efficiently transported into particle (i.e. electrons) acceleration in the downstream flow with the energy of $(\gamma-1) m_e c^2$, 
one can express the corresponding electron velocity ratio $\beta(\Theta_e)$ as 
\begin{eqnarray}
\beta(\Theta_e) \simeq \frac{\sqrt{\Theta_e (2+\Theta_e)}}{1+\Theta_e} \ ,
\end{eqnarray}
via the usual Lorentz factor $\gamma \equiv (1-\beta^2)^{-1/2}$. 
We consider that $\Theta_e(r)$ primarily characterizes the upper cut-off energy of the accelerated electron number distribution in the downstream flow, i.e., $\beta_2(r) = \beta(\Theta_e)$, whereas the lower cut-off is arbitrarily assumed to be $\beta \equiv \beta_1=0.01$ in this work. Hence, shock heating can produce a nonthermal electron distribution in the downstream flow in the form of the power-law; i.e. the electron spectrum distribution is assumed to obey $\epsilon^q$ \citep[e.g.][]{DroegeSchlickeiser86} as speculated in the solar flares. In our work we assume a conservative slope of $q=-2$ \citep[e.g.][]{ZW13} while the spectrum is only weakly sensitive to the exact value of $q$.   


In the conventional view of the standard  accretion disk scenario, disk surface radiates like blackbody of different temperature at a given point on the disk. The local intensity of disk blackbody between energy $\epsilon$ and $\epsilon+d\epsilon$  that is liberated at a point ($r, \phi$) on the disk surface is given by  Planck distribution  
\begin{eqnarray}
B_D(\epsilon, kT_{\rm in}) d\epsilon= \frac{8\pi \epsilon^2}{h^2 c^2} \frac{1}{e^{\epsilon/kT_{\rm in}}-1} d\epsilon  \ , \label{eq:voigt}
\end{eqnarray}
where $\epsilon$ is the thermal photon energy in a local disk frame and $kT_{\rm in}$ denotes the maximum disk temperature in the Shakura-Sunyaev model with GR correction as 
\begin{eqnarray}
kT_{\rm in} \approx 10 \left(\frac{\dot{m}}{0.5}\right)^{1/4} \left(\frac{1}{m_8}\right)^{1/4} \left(\frac{r}{r_g}\right)^{-3/4} \left\{1- \left(\frac{r_{\rm in}}{r} \right)^{1/2} \right\}^{1/4}~\textmd{eV} \ , \label{eq:Tin}
\end{eqnarray}
emitted at the characteristic radius $r= r_{\rm D} \equiv (49/6) r_g \approx 8 r_g$ for $a/M=0$ \citep[e.g.][]{Frank92,Kato08}. While in reality the disk radiation is known to be multi-color spectrum \citep[][]{Mitsuda84}, in this work we assume that the hottest part of the disk predominantly contributes to a subsequent Comptonization. In other words, seed disk photons for Comptonization are assumed to originate primarily from $r=r_D$ where the disk temperature is maximum. Note, however, that both $kT_{\rm in}$ and $r_D$ depend  on the BH spin $a$. 
The disk continuum  is then reprocessed by Compton up-scattering via energetic electrons 
in  the hot downstream region with a shock front at $r=r_{\rm sh}(< r_D)$ to produce the soft excess.

Following the same formalism by \citet{ZW13}, the differential Compton spectrum (i.e. Compton flux generated  per solid angle per radius) in the local reference frame of downstream flow can be described as photon spectrum  with $B_{\rm pl}(\epsilon,kT_{\rm in})$ being the blackbody intensity seen in the plasma frame; i.e. photon counts per time per area per energy  is expressed as
\begin{eqnarray}
I_{\rm Comp} \left(\epsilon',kT_{\rm in}\right) \propto \frac{1}{H_1 \left(\beta_1,\beta_2 \right)} \int_{\epsilon_o}^{\epsilon'} \frac{B_{\rm pl} \left(\epsilon_{\rm pl}, kT_{\rm in} \right)}{\epsilon_{\rm pl}}  H \left(\frac{\epsilon'}{\epsilon_{\rm pl}} \right) d\epsilon_{\rm pl}  \ , \label{eq:voigt}
\end{eqnarray}
where $\epsilon_{\rm pl}$ and $\epsilon'$ are respectively the incoming disk photon energy and the outgoing Comptonized photon energy measured in  the rest-frame of the downstream plasma and    
\begin{eqnarray}
H \left(\frac{\epsilon'}{\epsilon}\right) &=& 
 \left\{
\begin{array}{llr} \left. \int_{\beta_1}^{\beta_2} \right.
  \left. (\gamma-1)^{-q} \gamma^{-1} \beta^{-3} \zeta(\frac{\epsilon'}{\epsilon},\beta) \right. d\beta &
~~{\rm if}~~~ 1 \le \frac{\epsilon'}{\epsilon} < \frac{1+\beta_1}{1-\beta_1} ,  
\\
  \int_{\beta_c}^{\beta_2} 
  \left. (\gamma-1)^{-q} \gamma^{-1} \beta^{-3} \zeta(\frac{\epsilon'}{\epsilon},\beta) \right. d\beta & 
~~{\rm if}~~~ \frac{1+\beta_1}{1-\beta_1} \le \frac{\epsilon'}{\epsilon} \le \frac{1+\beta_2}{1-\beta_2}, 
\\
  0 & 
~~{\rm if}~~~ \frac{1+\beta_2}{1-\beta_2} < \frac{\epsilon'}{\epsilon} , 
\end{array} \right. \label{eq:column1}
\\
\zeta \left(\frac{\epsilon'}{\epsilon},\beta \right) &=& \frac{\epsilon}{\epsilon'} (\beta +1) \left(\beta ^2+1\right) \gamma ^2 - \frac{\epsilon'}{\epsilon} \frac{1
   }{(\beta +1) \gamma ^2} \nonumber \\
& & + 2 \left[ \ln \left\{ \frac{\epsilon'}{\epsilon} \frac{1}{(\beta +1)^2
   \gamma^2} \right\} +\frac{\beta }{\beta +1} \right] \ , \label{eq:voigt}
\end{eqnarray}
with a cut-off plasma velocity $\beta_c$ defined as
\begin{eqnarray}
\beta_c &\equiv& \frac{\epsilon'/\epsilon-1}{\epsilon'/\epsilon+1} \ ,   
\end{eqnarray}
and the lower integration limit $\epsilon_o$ is given by
\begin{eqnarray}
\epsilon_o (\epsilon', \beta_2) = \epsilon' \left(\frac{1-\beta_2}{1+\beta_2} \right) \ . \label{eq:voigt}
\end{eqnarray}
Lastly, the above coefficient $H_1(\beta_1,\beta_2)$ in equation~(16) has been defined as
\begin{eqnarray}
H_1 (\beta_1, \beta_2) \equiv \int_{\beta_1}^{\beta_2} \gamma^5 \beta \left(1+\frac{\beta^2}{3} \right)  (\gamma-1)^{-q}\ d\beta \ . \label{eq:voigt}
\end{eqnarray}

In addition, in terms of  the photon energy, it is crucial to take into account the redshift factor of photons as they propagate from one point to the other under the curved spacetime because the Comptonizing region -- hot downstream flow -- is in a close proximity to the BH. 
Realizing that the postshock downstream region in this scenario typically lies within the ISCO (for both spinning and non-spinning BH cases)  as depicted in Figure~\ref{fig:f1}, disk photons are strongly subject to relativistic effects in several ways while they propagate from the disk surface at $r=r_{\rm D}$ to the plasma downstream region at ($r,\phi$) where $r_H \le r \le r_{\rm sh} < r_D$ and $0 \le \phi \le 2\pi$; i.e. classical Doppler motion of the downstream plasma, special relativistic time dilation and gravitational redshift \citep[e.g.][]{Bardeen72,Cunningham75, Kojima91,Hollywood97}.
The photon frequency is then shifted between the disk frame of the Keplerian motion and accreting plasma by factor of 
\begin{eqnarray}
g_1(r, \phi) \equiv  \left| \frac{(p_\mu u^\mu)_{\rm pl}}{(p_\mu u^\mu)_{D}} \right|_{\rm Disk \rightarrow Plasma}   \ , \label{eq:g1}
\end{eqnarray}
where $p_\mu = (-E_{\rm ph}$ , $\pm E_{\rm ph} (R(r))^{1/2}/\Delta$, $\pm (\Theta(\theta))^{1/2}$, $\xi E_{\rm ph})$ denotes four-momentum of photons defined respectively in the plasma rest-frame (``pl") and the disk rest-frame (``D").   In equation~(22) above, note that we numerically solve null geodesic equations by GR ray-tracing (see \S 3.2) in radial $R(r)$ and angular $\Theta(\theta)$ directions with the photon energy $E_{\rm ph}$ in the emitted frame and its angular momentum component $\xi$ \citep[][p.347]{Chandra83}.     
The disk blackbody intensity in equation~(14)  in the plasma downstream frame  can then be expressed as %
\begin{eqnarray}
B_{\rm pl} (\epsilon_{\rm pl}, kT_{\rm in}) = g_1^3 B_D \left(\epsilon_D,kT_{\rm in} \right) \ ,\label{eq:voigt}
\end{eqnarray}
where $\epsilon_D \equiv \epsilon_{\rm pl}/g_1$ is the seed photon energy in the disk frame using  Lorentz invariance of photon intensity. Note that $g_1$ depends on $(r,\phi)$ because of relative motion between the Keplerian disk and accreting plasma. 
These photons are then locally Comptonized in all parts of the postshock flow as expressed in equations~(16)-(21).

Assuming that the Comptonization takes place spatially uniformly within the downstream region ($r_H \le r \le r_{\rm sh}$), the observed Comptonized intensity is given by
\begin{eqnarray}
I_{\rm obs} (\epsilon_{\rm obs},kT_{\rm in}) \propto  g_2(r,\phi)^3 I_{\rm Comp}(\epsilon',kT_{\rm in}) = g_2^3 I_{\rm Comp} \left(\frac{\epsilon_{\rm obs}}{g_2}, kT_{\rm in} \right) \ , \label{eq:voigt}
\end{eqnarray}
where 
\begin{eqnarray}
g_2(r, \phi) \equiv \left| \frac{(p_\mu u^\mu)_{\rm obs}}{(p_\mu u^\mu)_{\rm pl}} \right|_{\rm Plasma \rightarrow Observer} \ . \label{eq:g2}
\end{eqnarray}
provides the additional redshift factor of the reprocessed photon energy due to secondary  relativistic effects with respect to a distant observer  while those photons propagate under curved spacetime from the downstream plasma to the observer, as illustrated in Figure~\ref{fig:f1}, such that $\epsilon' \equiv \epsilon_{\rm obs}/g_2$. 
Therefore, the observed Comptonized  intensity per radius is  affected by the coupling between the two redshift effects, $g_1$ and $g_2$, and found by 
\begin{eqnarray}
I_{\rm obs} (\epsilon_{\rm obs},kT_{\rm in}) \propto  \frac{g_1(r,\phi)^3 g_2(r,\phi)^3}{H_1(\beta_1,\beta_2)} \int_{}^{} \frac{B_D(\epsilon_{\rm pl}/g_1, kT_{\rm in})}{\epsilon_{\rm pl}} d \epsilon_{\rm pl} \ , \label{eq:voigt}
\end{eqnarray}
Note, in equations~(22) and (25), that different photon trajectories are considered and therefore $g_1$ is totally independent of $g_2$. 
Thus, the corresponding differential flux via Comptonization is given by integrating over a solid angle subtended by the downstream region as
\begin{eqnarray}
\frac{dF_{\rm obs}}{dr} \propto \int\int_{\rm downstream} I_{\rm obs}\left(\epsilon_{\rm obs},kT_{\rm in} \right) d\Omega_{\rm obs}  \ . \label{eq:voigt}
\end{eqnarray}
By further integrating over the downstream region in radius (i.e. $r_H \le r \le r_{\rm sh}$) one can calculate the observed Comptonized spectrum 
\begin{eqnarray}
F_{\rm obs} = \int_{r_H}^{r_{\rm sh}} \left( \frac{dF_{\rm obs}}{dr} \right) dr   \ , \label{eq:voigt}
\end{eqnarray}
where $r_H \equiv [ 1+(1-a^2/M^2)^{1/2} ] r_g$ is the radius of the event horizon normalized by the gravitational radius $r_g$ and $r_{\rm sh}$ is the radius of the shock front in the accreting plasma.  Lastly, the normalization of the model spectrum is currently treated as an independent parameter to be constrained by data (see \S 5). 

\begin{figure}[ht]
\begin{center}$
\begin{array}{cc}
\includegraphics[trim=0in 0in 0in
0in,keepaspectratio=false,width=3.0in,angle=-0,clip=false]{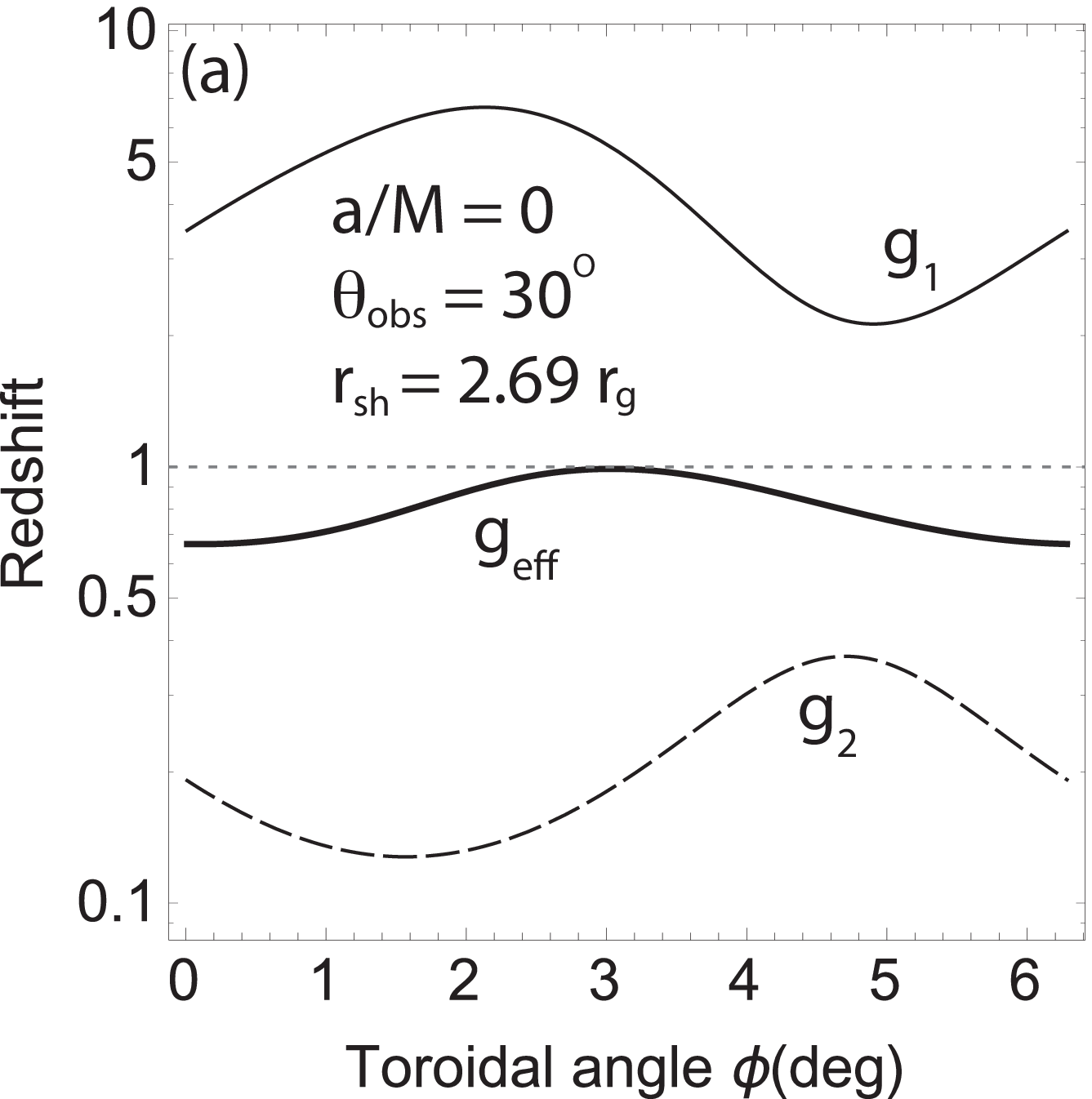} &
\includegraphics[trim=0in 0in 0in
0in,keepaspectratio=false,width=3.0in,angle=-0,clip=false]{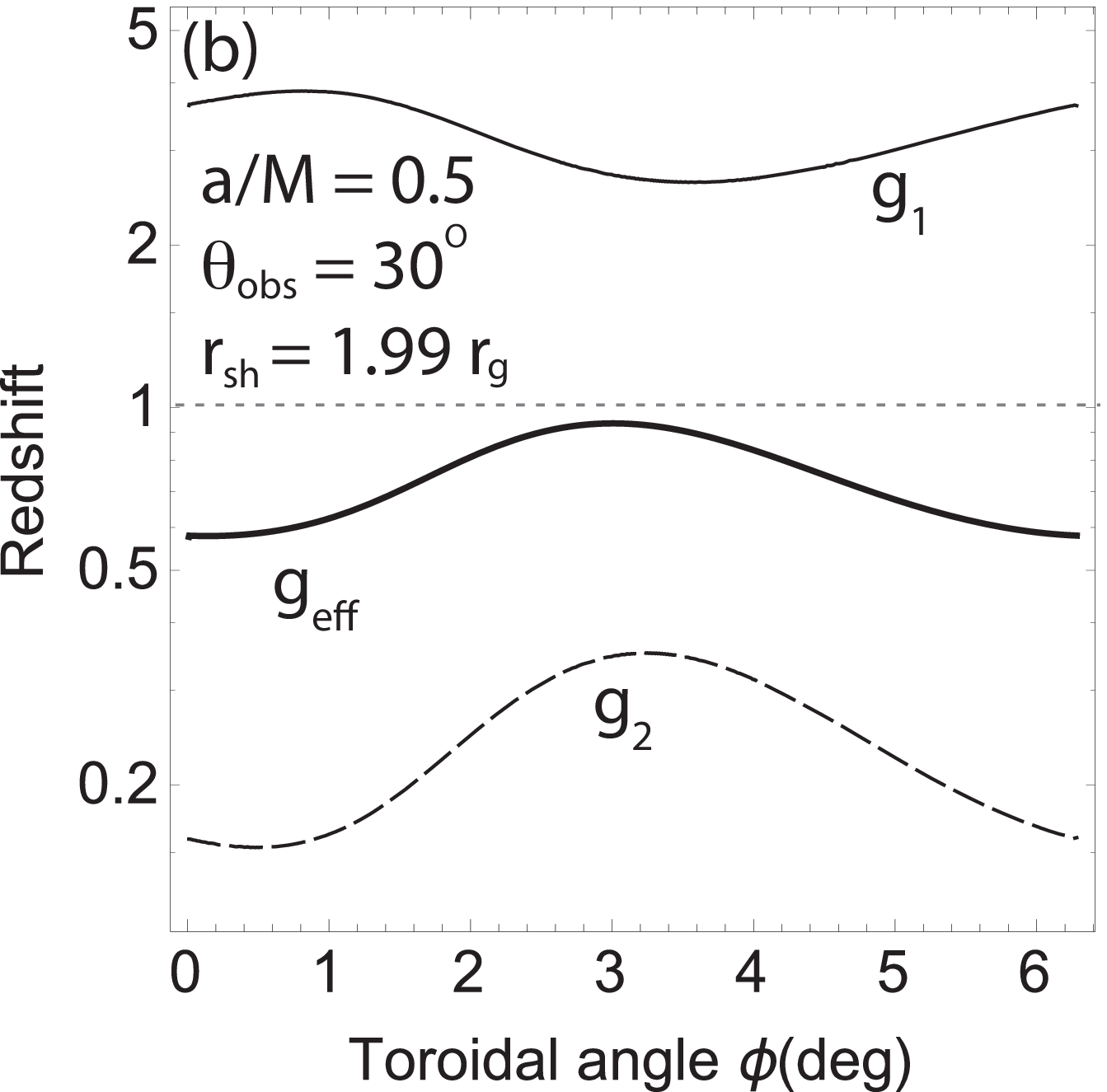} 
\end{array}$
\end{center}
\caption{The transfer functions, $g_1(r_{\rm sh}, \phi)$, $g_2(r_{\rm sh}, \phi)$ and $g_{\rm eff}(r_{\rm sh},\phi) \equiv g_1 \cdot g_2$, for the fiducial plasma accretion with (a) $a/M=0$  and (b) $0.5$  as a function of  toroidal angle $0 \le \phi \le 2 \pi$ at the shock location $r=r_{\rm sh}$. See texts for the definitions and Table~\ref{tab:tab1} for detailed parameters. } \label{fig:f6}
\end{figure}

\subsection{Transfer Functions with GR Ray-Tracing}

Given a kinematic field of MHD accretion coupled with the conventional thermal disk radiation, one can calculate two redshift factors expressed in Eqns.~(\ref{eq:g1}) and (\ref{eq:g2}).
%
%
It is reminded that these factors are functions of radius and toroidal positions of plasma and the observed photon flux also depends on both $g_1$ and $g_2$ due to relativistic beaming. 
While incoming disk photons traveling towards the downstream plasma are always blueshifted ($g_1>1$), Comptonized photons from the plasma region are always redshifted ($g_2<1$)  due to strong gravitational redshift being dominant over the longitudinal Doppler effect (at least for small angle $\theta \le 45\degr$ considered here). The transfer function (i.e. effective redshift factor), $g_{\rm eff}$, is then the product of the two, $g_{\rm eff} \equiv g_1 g_2$.  

Employing the standard GR ray-tracing approach we calculate and store the transfer functions for different sets of inclination angle $\theta$, BH spin $a$ and the fiducial accretion solutions. 
As an example, Figure~\ref{fig:f6} shows  $g_1, g_2$ and $g_{\rm eff}$ as a function of azimuthal angle $0 \le \phi \le 2 \pi$ at the shock location $r=r_{\rm sh}$ for (a) $a/M=0$  and (b) $a/M=0.5$ assuming $\theta=30\deg$ with fiducial shocked plasma solutions similar to those in Table~\ref{tab:tab1}. 
With respect to the initial disk photon energy, the observed photons are found, as expected, to be always redshifted despite the blueshift from $g_1$ because the Comptonizing region (i.e. downstream plasma) is sufficiently close to the horizon allowing for the factor $g_2$ to always dominate over the factor $g_1$ in redshift such that $0.5 \lesssim g_{\rm eff} \lesssim 1.0$ as calculated. Note that the range of the effective redshift factor for both $a/M=0$ and $0.5$ appears to be very similar despite the spatial difference in the downstream plasma region; i.e. closer in towards the horizon for $a/M=0.5$ case. This is due to a competition  between $g_1$ and $g_2$ for a given BH spin. In other words, $g_1$ may be comparatively very large for $a/M=0.5$ case because the downstream region is further closer in, while $g_2$ should then be correspondingly very small as Comptonized photons must climb up farther out towards the observer. As a result, a modest value of the effective redshift $g_{\rm eff}$ is always achieved almost regardless of the BH spin. 
These profiles are also found to be almost independent of radius $r_H \le r \le r_{\rm sh}$ since the downstream region is very compact in radius.    
In \S 3.3 the computed transfer function is then used to calculate the Comptonized spectrum in combination with the MHD plasma solutions.    


\vspace{-0in}
\begin{deluxetable}{l|c}
\tabletypesize{\small} \tablecaption{Grid of {\tt compsh} Model Parameters } \tablewidth{0pt}
\tablehead{Primary Parameter & Value  }
\startdata
BH Spin $a/m$ & $-0.5, 0, 0.5$ \\
Inclination Angle $\theta$ (degrees) & $15\deg, 30\deg, 45\deg$ \\
Disk Temperature $kT_{\rm in}$ (eV) & 10, 20, 30, 40 \\
Downstream Elecetron Energy $kT_e$ & See \S 3.3 \\
\enddata
\vspace{0.05in}
\label{tab:tab2}
\end{deluxetable}

\clearpage

\subsection{Modeling Comptonized Spectra  }

Based on the numerical approach described in \S 3.1 and 3.2 we calculate the Comptonized spectra for various shocked plasma accretion with sets of fiducial disk blackbody temperature. Although in principle the effective disk temperature $kT_{\rm in}$ is strictly determined by the BH spin $a$, it is often speculated that the actual disk radiation is most likely subject to various scattering and reflection due to the (presumably corona-related) atmospheric property above the disk, which could slightly alter (if not substantially) the effective temperature of emerging thermal radiation (aka. color temperature) making the degree of color temperature index $f_c$ be a sensitive quantity to assess the spectral property of the local thermal radiation. Following a number of previous works on disk color temperature \citep[e.g.][]{Ross92,ShimuraTakahara95,Li05,Davis05,Done12}, we adopt the conventional value of $f_c=1.7$ in this paper. Because of this uncertainty in estimating  the exact disk temperature,  in our calculations we introduce $kT_{\rm in}$ replacing the conventional $kT_{\rm in}$ and treated as a free parameter in the likelihood of the expected value (i.e. $10-40$ eV) for a typical AGN disk environment. 
Among other independent model parameters, the primary ones to be intensively explored in this work include downstream electron energy\footnote[4]{Downstream electron energy $kT_e$ is actually a dependent parameter determined by a shock location $r_{\rm sh}$ for a given set of conserved plasma quantities. Nonetheless, we employ this variable in the current scheme to construct a library of model spectra later in \S 4.} $kT_{e}$, disk temperature $kT_{\rm in}$ and inclination angle $\theta_{\rm obs}$ as listed in Table~\ref{tab:tab2}. Hence, the entire model spectrum in the present approach is characterized by these {\it three} quantities. By seeking a best-fit model, one can constrain the three parameters followed by the other important plasma quantities.  

\begin{figure}[t]
\begin{center}$
\begin{array}{cc}
\includegraphics[trim=0in 0in 0in
0in,keepaspectratio=false,width=3in,angle=-0,clip=false]{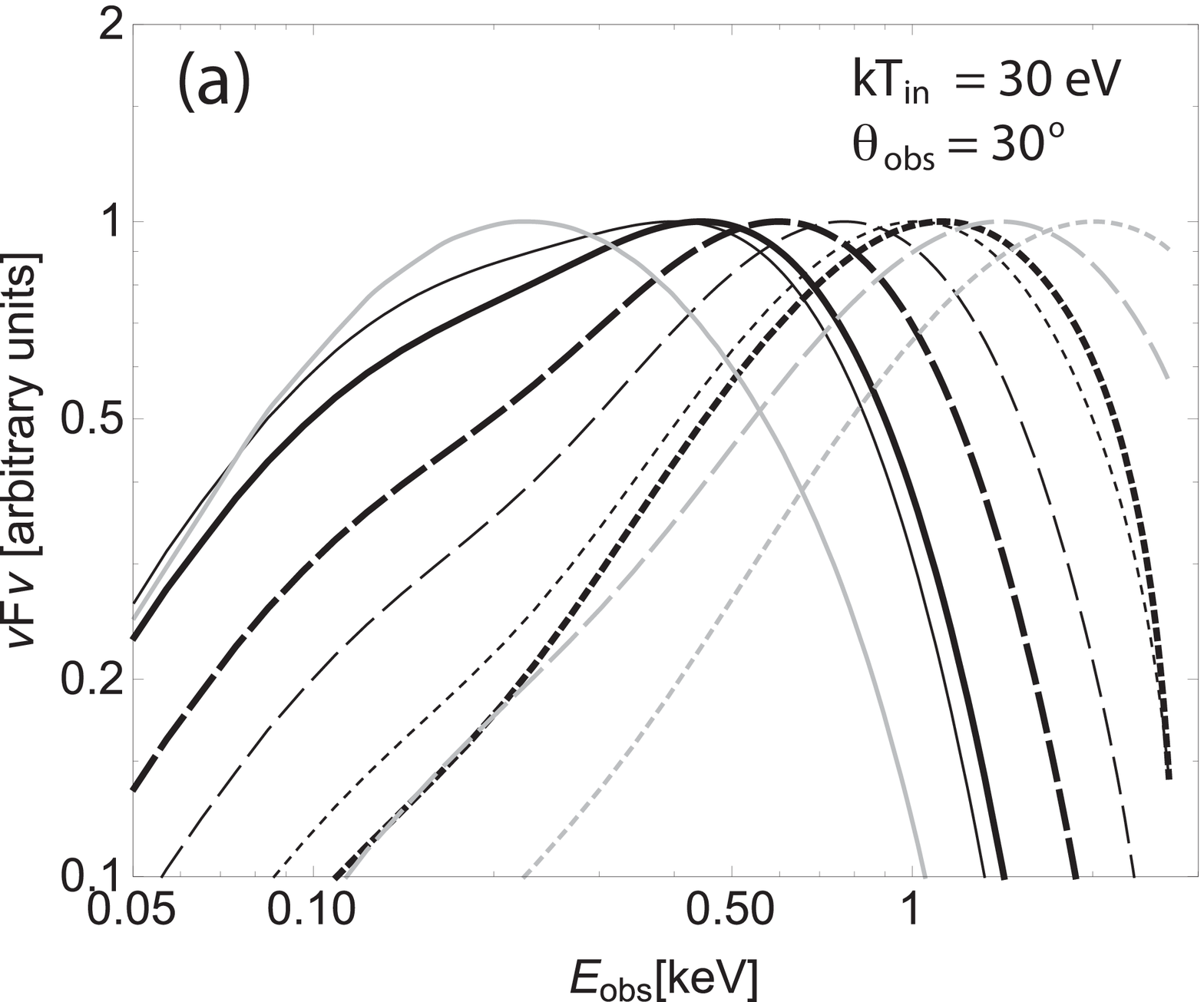} &
\includegraphics[trim=0in 0in 0in
0in,keepaspectratio=false,width=3in,angle=-0,clip=false]{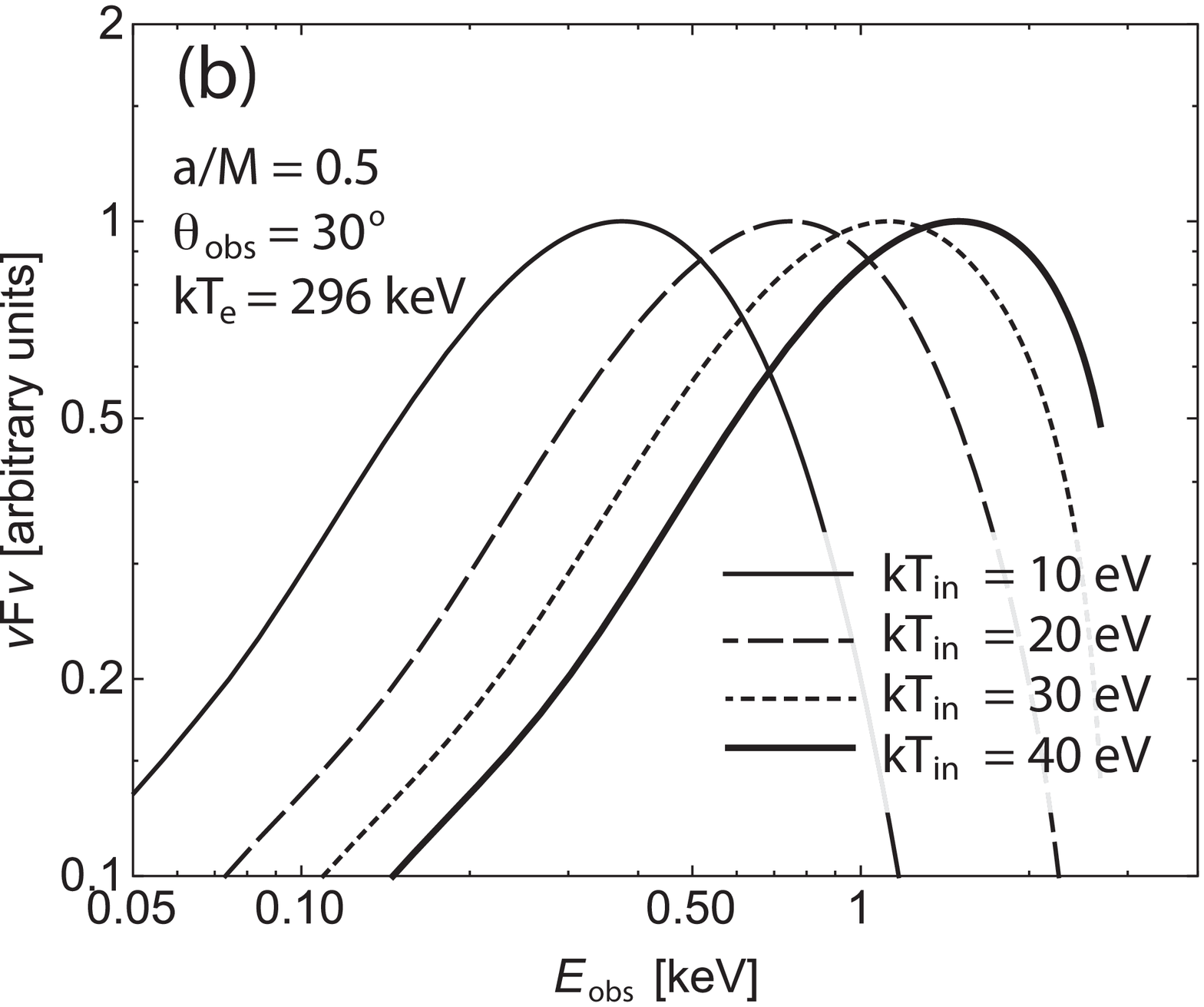} 
\end{array}$
\end{center}
\caption{(a) Normalized Comptonized spectra $\nu F_\nu$ in the observer's frame (without cosmological redshift) for various electron energy $kT_e$ with $kT_{\rm in}=30$ eV and $\theta=30\deg$: $kT_{e}=33$ keV, $250$ keV and $378$ keV for $a/M=-0.5$ (gray), $75$ keV, $125$ keV and $256$ keV for $a/M=0$ (dark), and $126$ keV, $179$ keV and $296$ keV for $a/M=0.5$ (thick solid). In each BH spin, temperature is lowest in {\it solid} curve, intermediate in {\it dashed} curve and highest in {\it dotted} one. (b) Normalized Comptonized spectra for various disk temperature  with the fiducial accretion flow of $a/M=0.5$, $kT_e=296$ keV and $\theta_{\rm obs}=30\deg$ (see Table~\ref{tab:tab1}): we show $kT_{\rm in}=10, 20, 30$ and $40$ eV.  } \label{fig:f7}
\end{figure}

Figure~\ref{fig:f7}a presents the calculated normalized spectra (in $\nu F_\nu$) for different downstream electron energy $kT_{e}$ assuming $kT_{\rm in}=30$ eV and $\theta_{\rm obs}=30\degr$ for a given different BH spin. In each spin value, three spectra are computed with the lowest (solid), intermediate (dashed) and highest (dotted) shocked electron energy with $kT_{\rm in}=30$ eV and $\theta_{\rm obs}=30\deg$; we select $kT_e=33$ keV, $250$ keV and $378$ keV for $a/M=-0.5$ (gray); $kT_e=75$ keV, $125$ keV and $256$ keV for $a/M=0$ (solid); $kT_e=126$ keV, $179$ keV and $296$ keV for $a/M=0.5$ (thick). 
We stress here again that the energy $kT_e$ is a dependent quantity determined by the shock location in the model for a given BH spin, thus not arbitrarily selected a priori.
It is noted that the spectral peak can exceed $\sim 1$ keV depending on how much the downstream plasma is heated by the shock, and, as expected, they seem to be well correlated. The spectral shape is found to be more or less self-similar in a qualitative manner for different plasma flows and even BH spin values. This seems to be consistent with the observational fact that the detected soft excess can be almost uniquely accounted for in all AGNs with a single blackbody component of more or less the same temperature $kT \sim 0.1-0.2$ keV despite their potentially diverse circumnuclear conditions. 
Figure~\ref{fig:f7}b shows a similar calculation but for different disk temperature $kT_{\rm in}$ with $a/M=0.5$ and $\theta_{\rm obs}=30\deg$: $10$ eV (solid), $20$ eV (dashed), $30$ eV (dotted) and $40$ eV (thick) for the same photon emitting radius $r_D$. It is clear that the peak energy of the  emergent spectrum has a strong dependence on the disk temperature $kT_{\rm in}$ as expected. The expected Comptonized flux in this model is therefore strongly correlated with UV flux.   


\begin{figure}[ht]
\begin{center}$
\begin{array}{cc}
\includegraphics[trim=0in 0in 0in
0in,keepaspectratio=false,width=3in,angle=-0,clip=false]{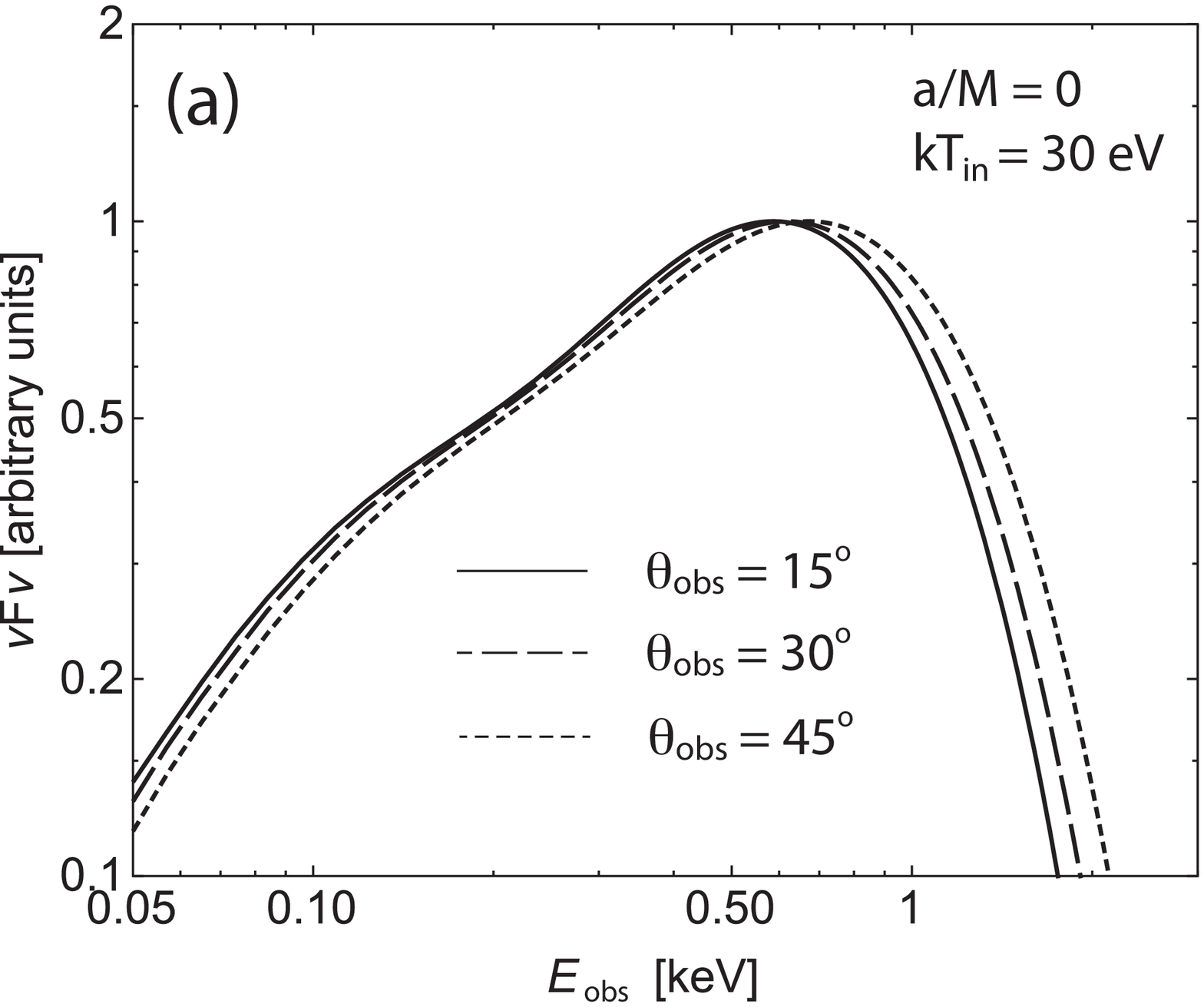} &
\includegraphics[trim=0in 0in 0in
0in,keepaspectratio=false,width=3in,angle=-0,clip=false]{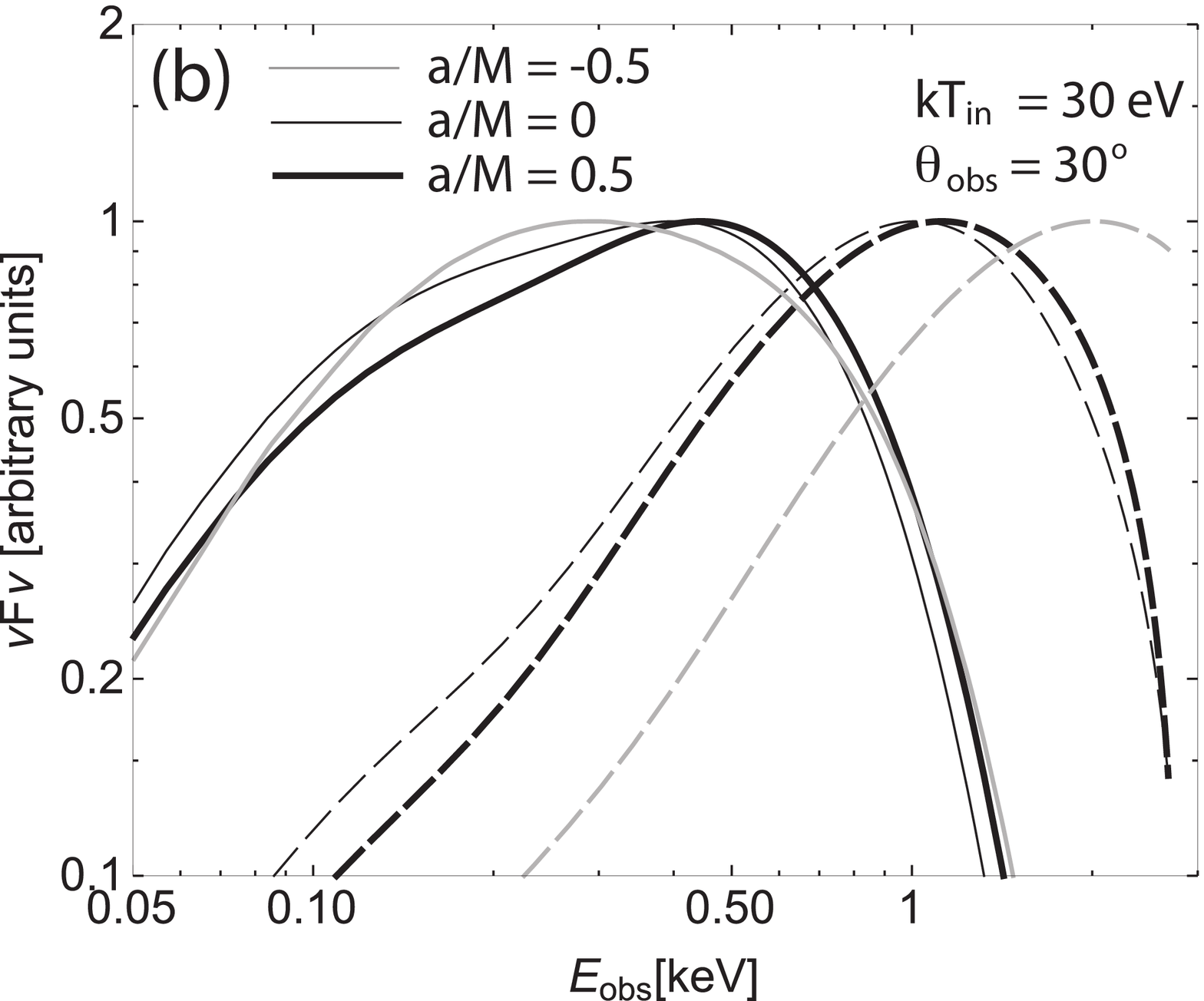} 
\end{array}$
\end{center}
\caption{(a) Normalized Comptonized spectra for various inclination $\theta$ with $a/M=0.5$, $kT_e=179$ keV and $kT_{\rm in}=30$ eV (see Table~\ref{tab:tab1}): we show $\theta_{\rm obs}=15\deg, 30\deg$ and $45\deg$. (b) Range of the normalized Comptonized spectra for various BH spin with $kT_{\rm in}=30$ eV and $\theta=30\deg$: solid/dashed curves are obtained from the lowest/highest electron energy considered for BH spin $a/M=-0.5$ (gray),  $0$ (dark) and $0.5$ (thick solid).} \label{fig:f8}
\end{figure}

We also consider the effect of different inclination angle in Figure~\ref{fig:f8}a for $\theta_{\rm obs}=15\deg$ (solid), $30\deg$ (dashed) and $45\deg$ (dotted) assuming $a/M=0$ and $kT_{\rm in}=30$ eV. We find that there is little change in the spectrum due to viewing angle. This is because the photon emitting region (i.e. downstream plasma) is generically (i) very close to the BH where gravitational redshift is predominant and (ii) the size of hot region is also very narrow in radius.  Hence, in such proximity to the black hole  the longitudinal Doppler blueshift never becomes significant enough to overcome the degree of gravitational redshift in all three cases. 
For this reason, this weak angle-dependence is seemingly very different from what one typically finds in the disk emission line, for example, broad iron fluorescence \citep[e.g.][]{Fabian89,Kojima91}.   A spectral variation due to different BH spin $a$ is examined in Figure~\ref{fig:f8}b where $a/M=0$ (dark), $0.5$ (thick) and $-0.5$ (gray) for $kT_{\rm in}=30$ eV and $\theta_{\rm obs}=30\deg$. Solid and dashed curve are obtained with the lowest and highest shocked electron energy in each BH spin, respectively.

As seen in this section, our calculations show that Comptonized disk photons in this scenario seem to be observationally plausible to account for the known soft excess feature in many Seyfert 1 X-ray spectra. To demonstrate its validity, we will apply this model in \S 4 to one of the well-known radio-quiet Seyfert 1 galaxy, Ark~120.

\section{Case Study: Ark~120 }

Based on the model spectra for the soft excess component as shown in \S 3, we now apply the model to Ark~120, one of the well-studied luminous Seyfert 1 AGNs ($z=0.0323$ and $M \approx 2 \times 10^8 \Msun$; e.g., \citealt{WandelPetersonMalkan99,Peterson04}) in which little absorption features have been detected in both UV and X-ray bands despite the persistent presence of the soft excess \citep[][]{TurnerPounds89,Brandt93} which makes this particular AGN a ``bare" nucleus \citep[e.g.][]{Vaughan04}.

As a case study to conclude the current work, we analyze a 60 ks {\it XMM-Newton}/EPIC-pn spectrum of Ark~120 observed in 2003 with the standard {\tt XSPEC} 12.8.2 package  \citep{Arnaud96} 
%
%
to perform the $\chi^2$-statistics. 
All fit parameters are given in the source rest-frame $(z = 0.0323$) and errors are quoted at the 90\% confidence level for one interesting parameter (i.e. $\Delta \chi^2 = 2.7$) unless otherwise stated. The Galactic column density, $N_H$, toward Ark~120 is fixed at $1.31 \times  10^{20}$ cm$^{-2}$ \citep[][]{Stark92,Dickey90}. Throughout this paper, $H_o = 70$ km s$^{-1}$ Mpc$^{-1}$ is assumed.

\begin{figure}[t]
\begin{center}$
\begin{array}{cc}
\includegraphics[trim=-0in 0in 0in
0in,keepaspectratio=false,width=3.0in,angle=-0,clip=false]{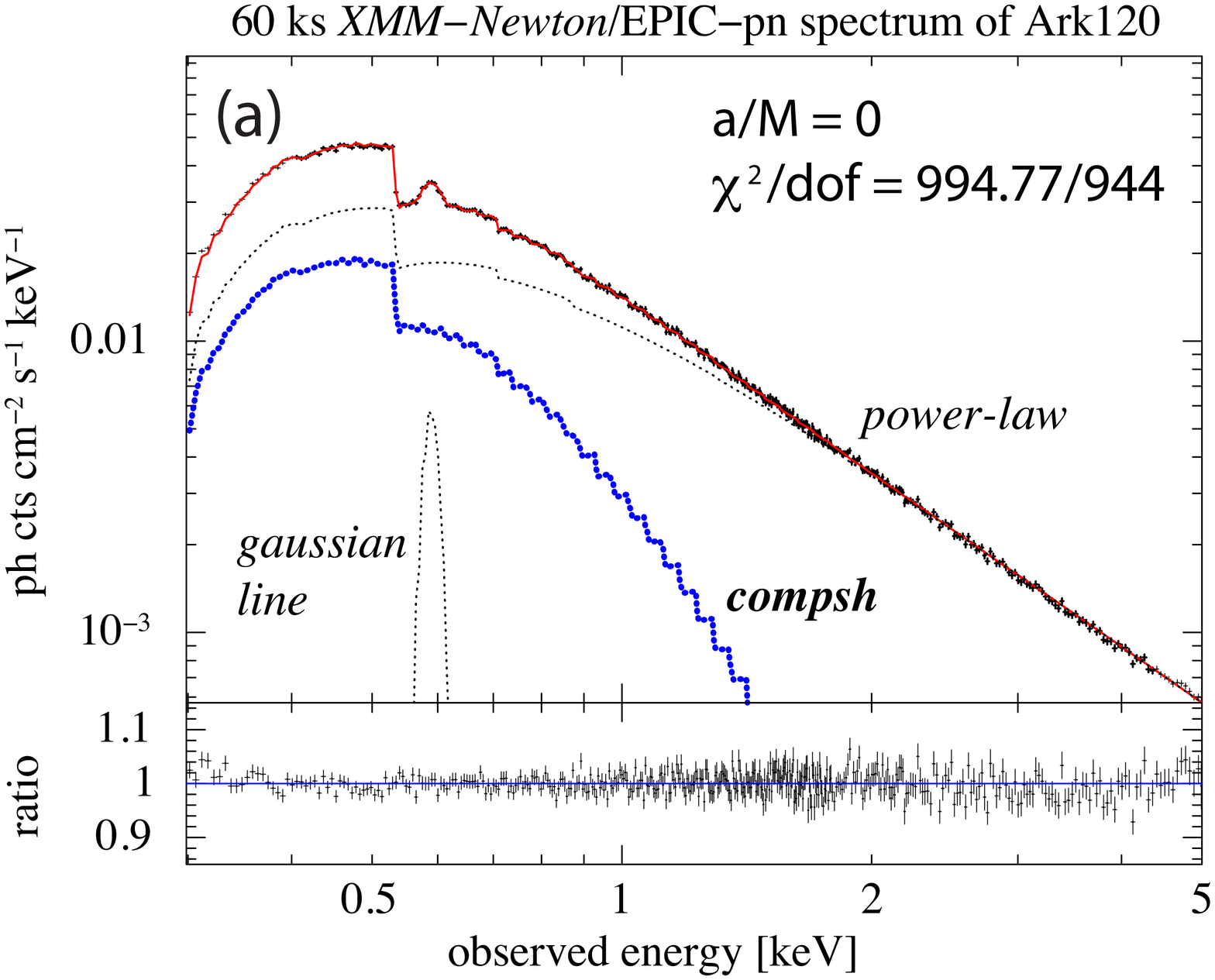} &
\includegraphics[trim=0in 0in 0in
0in,keepaspectratio=false,width=3.1in,angle=-0,clip=false]
{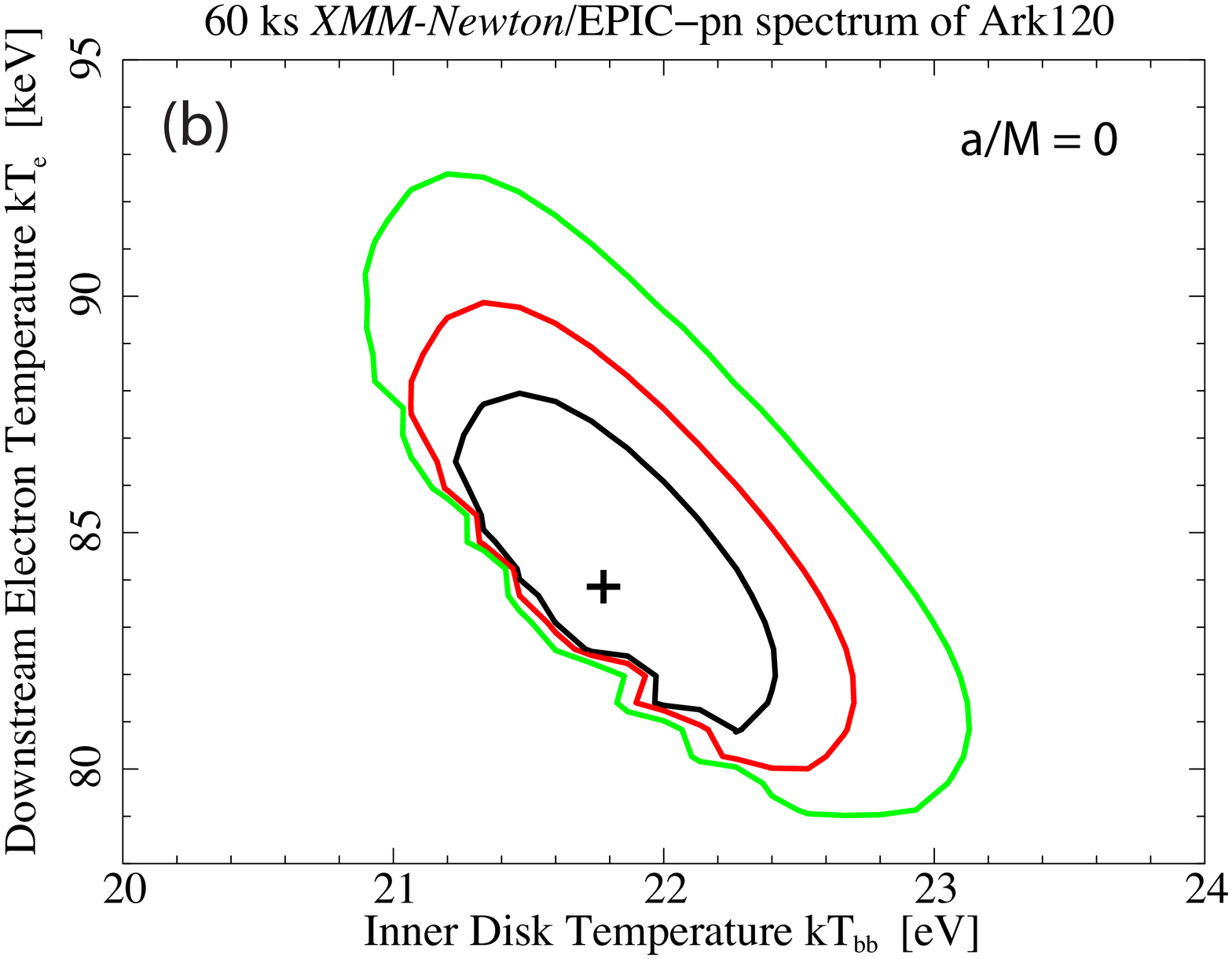}
\end{array}$
\end{center}
\caption{(a) 60ks {\it XMM-Newton}/EPIC-pn spectrum of Ark~120 fitted with the best-fit {\tt compsh} model for a \sw BH of $a/M=0$   and (b) its confidence contour plot showing 68\%, 90\%, and 99\% regions relative to the best-fit model. See Table~\ref{tab:tab3} for details. } \label{fig:f9}
\end{figure}

\subsection{Best-fit {\tt compsh} Model}

By constructing a grid of Comptonized spectra spanned by the model parameters listed in Table~\ref{tab:tab2}, we develop a {\tt compsh} model as an additive table model in {\tt XSPEC} tool whose free parameters are (i) effective disk photon temperature $kT_{\rm in}$ [eV], (ii) downstream electron energy $kT_e$ [keV], (iii) inclination angle $\theta_{\rm obs}$ [deg] and (iv) corresponding normalization. Following the analysis by \citet{Vaughan04} we freeze a single power-law of photon index $\Gamma=2$ to account for the continuum to the energy up to $5$ keV. 
As previously reported \citep[][]{Vaughan04,Matt14}, there has been  indicative of an additional component at $E \gtrsim 7-8$ keV due presumably to disk (blurred) reflection  as well as the well-defined iron emission line at $\sim 6.4$ keV. In this work  we don't consider a putative  reflection  component  since the proposed Comptonized model in this scenario is  not directly (if not completely) related to those hard X-ray photon production\footnote[5]{While reflection may be remotely related to disk photons via coronae, emitting regions in this scenario are physically very distant from each other and thus such a weak correlation, if any, would be smeared out.  }. 

%
After applying the {\tt compsh} model, we note a residual bump at $\sim 0.55$ keV  which can be attributed to the instrumental and Galactic oxygen edges \citep[][]{Vaughan04}. We independently treat this feature with a single gaussian line ({\tt zga}) as performed in \cite{Vaughan04}. Thus, our composite model is symbolically expressed  as {\tt phabs*(po+atable\{compsh\}+zga)} where {\tt po} is the power-law continuum and {\tt phabs} denotes the Galactic absorption. As stated in \S 3.3, the {\tt compsh} spectrum model is determined only by the above {\it three} primary parameters besides its normalization; i.e. $kT_{\rm in}$, $kT_e$ and $\theta_{\rm obs}$ and  independent of the other spectral components at least explicitly.
%

\begin{figure}[ht]
\begin{center}$
\begin{array}{cc}
\includegraphics[trim=-0in 0in 0in
0in,keepaspectratio=false,width=3.0in,angle=-0,clip=false]{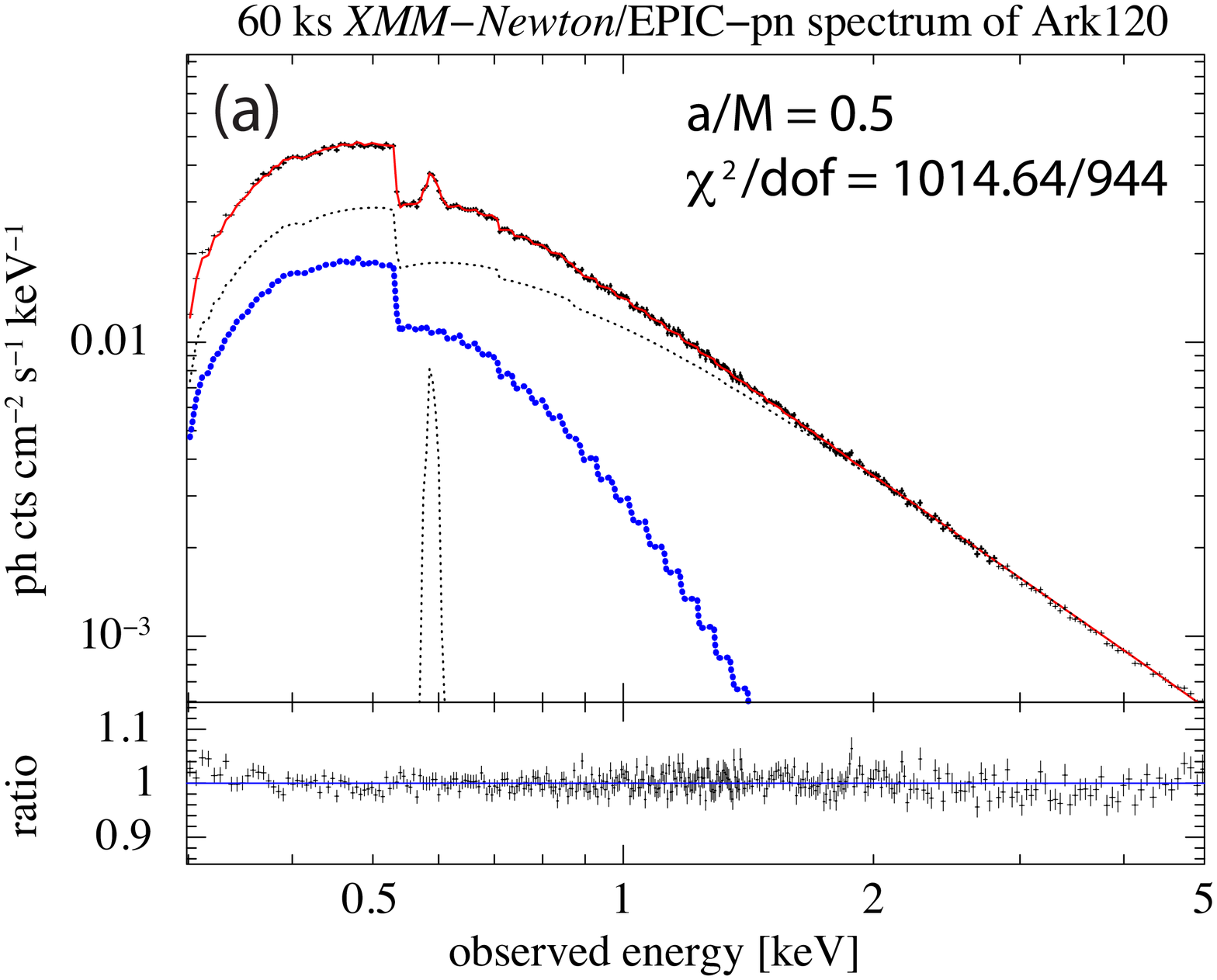} &
\includegraphics[trim=0in 0in 0in
0in,keepaspectratio=false,width=3.1in,angle=-0,clip=false]{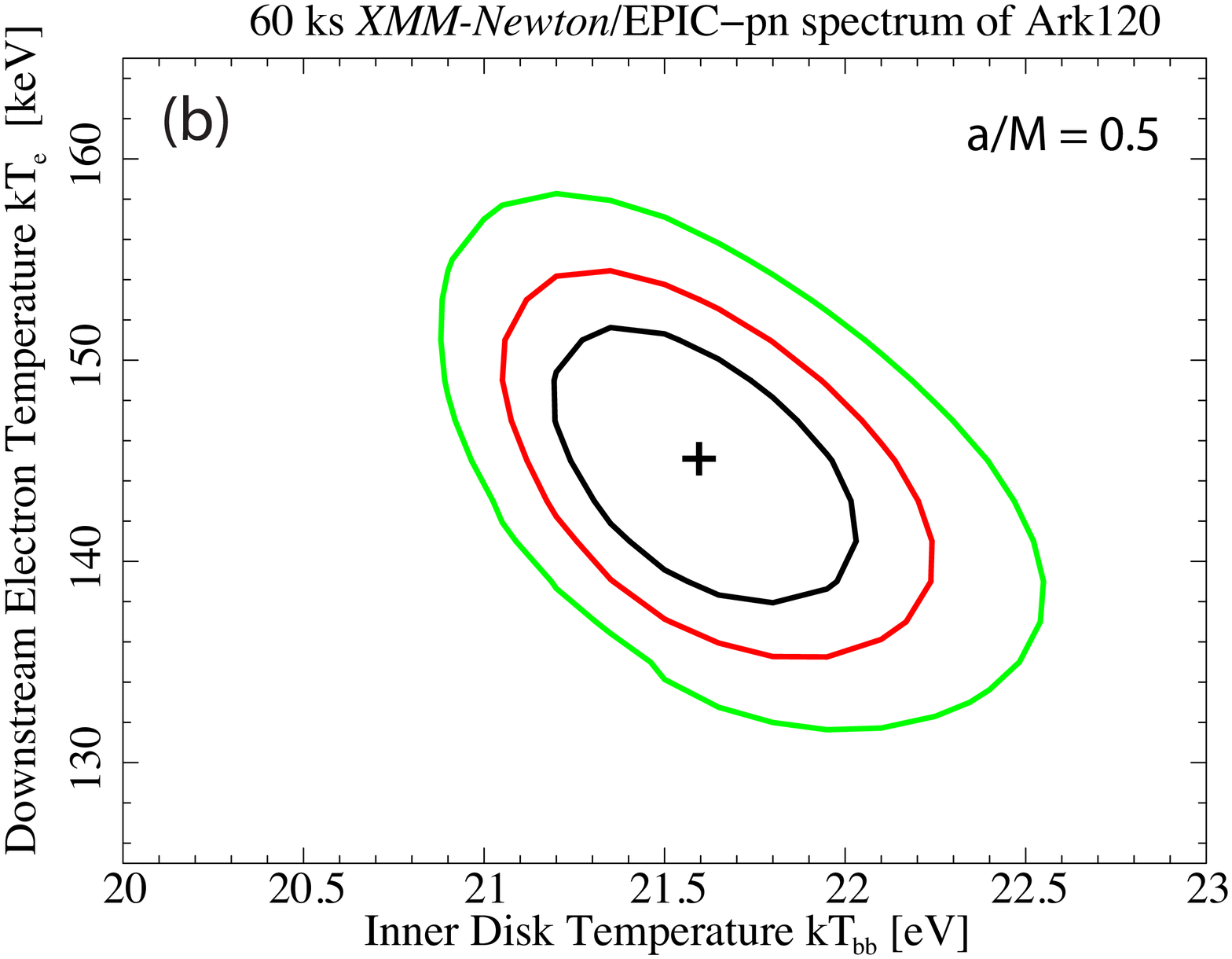}
\end{array}$
\end{center}
\caption{(a) Same as Figure~\ref{fig:f9} but for $a/M=0.5$ (prograde).} \label{fig:f10}
\end{figure}

\begin{figure}[ht]
\begin{center}$
\begin{array}{cc}
\includegraphics[trim=-0in 0in 0in
0in,keepaspectratio=false,width=3.0in,angle=-0,clip=false]{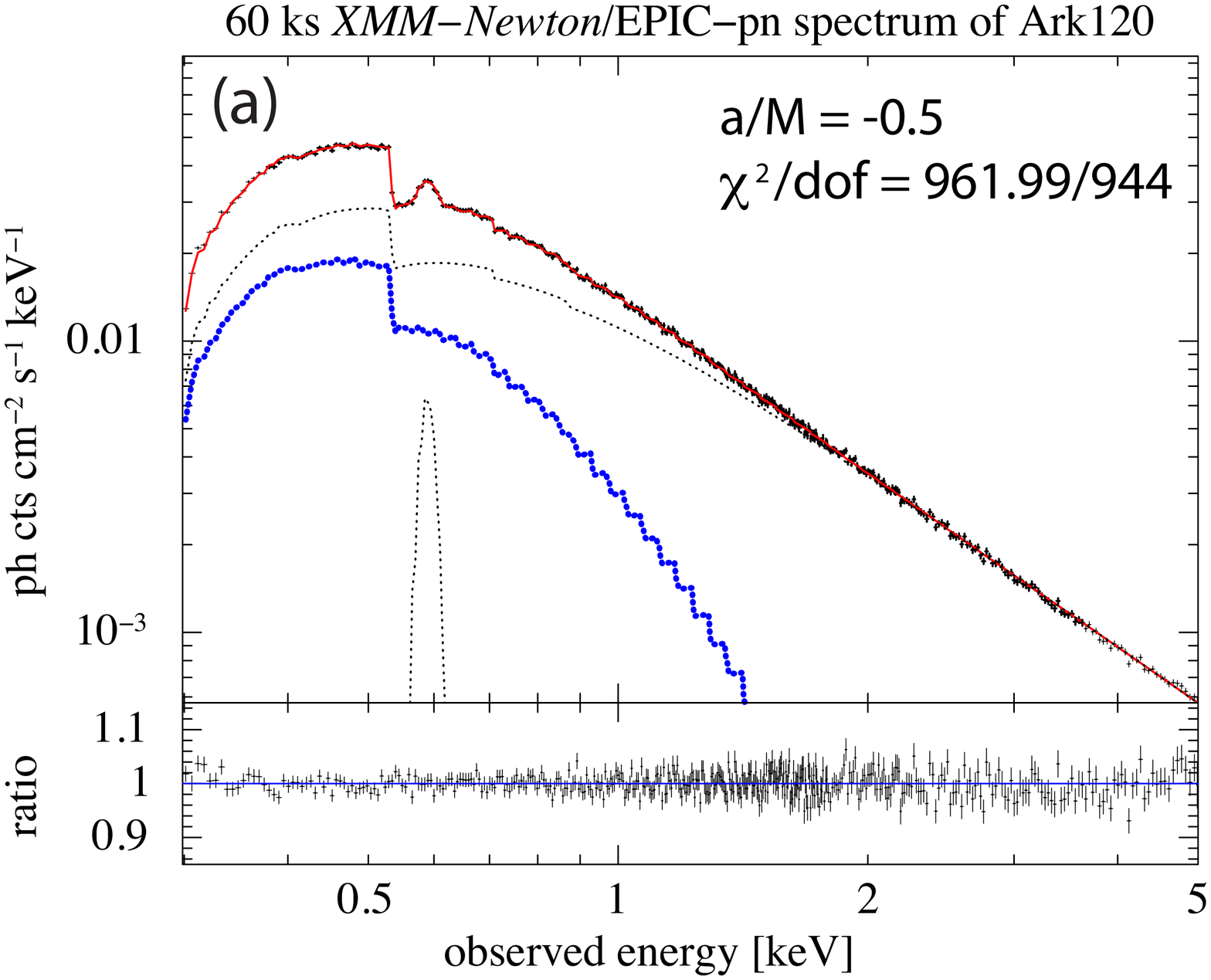} &
\includegraphics[trim=0in 0in 0in
0in,keepaspectratio=false,width=3.1in,angle=-0,clip=false]{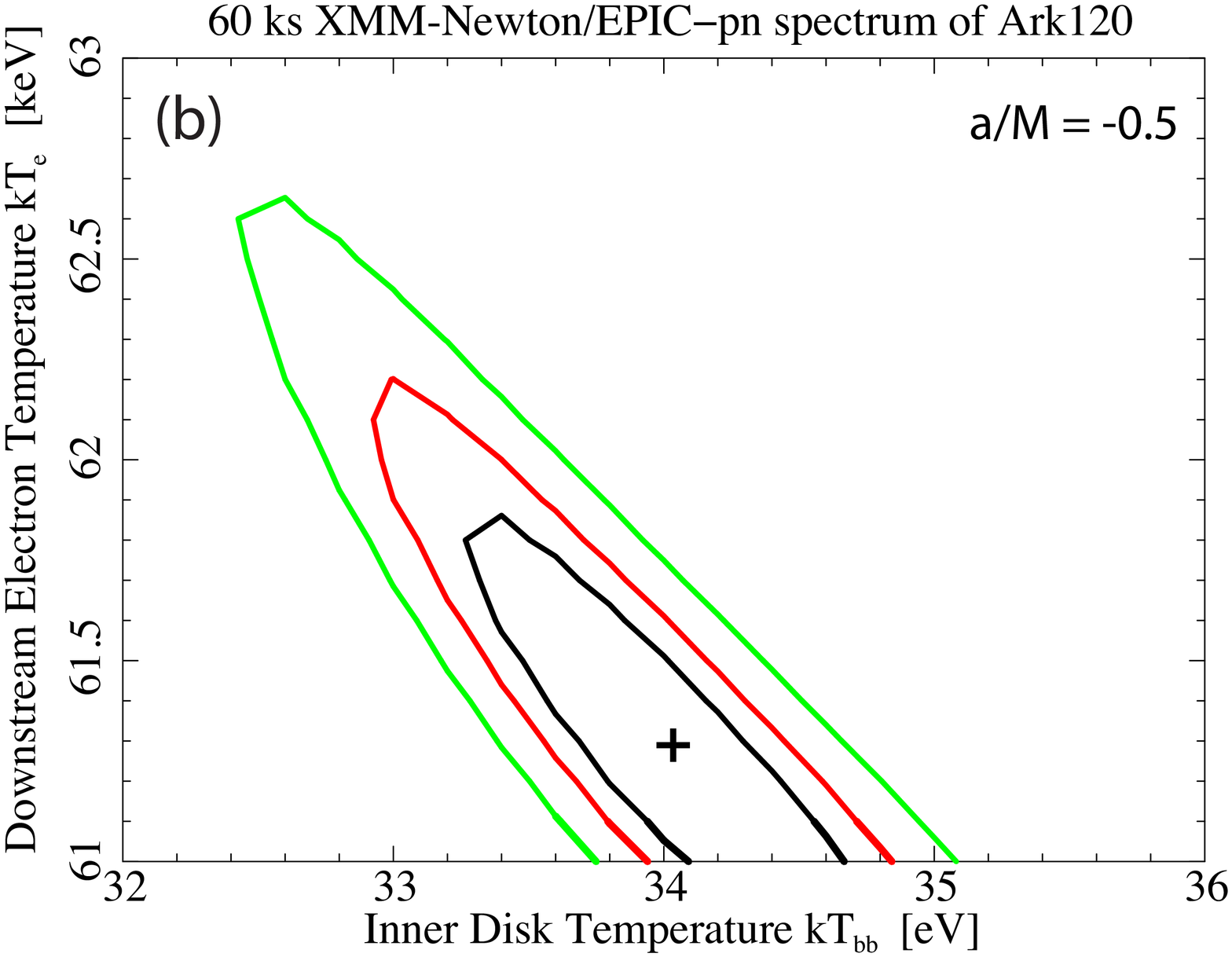}
\end{array}$
\end{center}
\caption{(a) Same as Figure~\ref{fig:f9} but for $a/M=-0.5$ (retrograde). } \label{fig:f11}
\end{figure}

We  present our results for $a/M=0$ (Fig.~\ref{fig:f9}), $a/M=0.5$ (Fig.~\ref{fig:f10}) and $a/M=-0.5$ (Fig.~\ref{fig:f11}) where a 60-ks {\it XMM-Newton}/EPIC-pn data is fitted with the {\tt compsh} model in each case. First, the soft excess is found to be successfully accounted for by the proposed model yielding an excellent statistical significance in which the derived best-fit parameters  are well constrained as shown in the contour plots at $68\%, 90\%$ and $99\%$ level as shown. 

The derived values of the best-fit parameters of the proposed models are listed in Table~\ref{tab:tab3}  for each case including the characteristic radii for plasma flows as well as the standing shock properties. Note that these radii derived in the table are not free-parameters but dependent variables.  We first note that all three cases are equally well constrained yielding a reasonable $\chi^2$ values for all three cases. With the seed disk photons of characteristic temperature ($\lesssim 20-30$ eV), the best-fit  downstream energy  $kT_e$ tends to increase with the BH spin primarily because the shock location tends to slightly shift inward  bringing the downstream region inward with the spin. The Comptonized spectra for $a/M=0.5$ are therefore subject to a more drastic gravitational redshift in the observer's frame requiring the downstream electron temperature $kT_e$ to be higher in the plasma rest-frame (before being redshifted) to balance. The derived viewing angle $\theta_{\rm obs}$ seems to be systematically low to intermediate for three cases consistent with the conventional classification scheme of the Seyfert galaxies such as Ark~120. Note also in Table~\ref{tab:tab3} that the characteristic magnetosonic radii and the \Alfven radius are all closer together in the case of $a/M=0.5$. The shock compression (i.e. shock strength), $n_2/n_1$, is well correlated with the electron energy $kT_e$ as compression is the source of generating additional  entropy in the downstream flow.  Since we are focusing in this work on fast MHD shocks, the field line becomes more refracted away from the shock normal (i.e. away from radial direction) across the shock front (i.e. $|B_{\phi,2}| > |B_{\phi,1}|$), the dissipated plasma energy is partially transformed to the field energy  consistent with increasing magnetization parameter $\sigma_2 > \sigma_1$ as shown in Table~\ref{tab:tab3}. The rate of increase in magnetization $\sigma$ due to shock is also well correlated with the BH spin (see, e.g., T02).  Among the three BH spins,  for $a/M=0$ and $0.5$ cases, the derived disk temperature is  statistically consistent ($kT_{\rm in} \approx 21-23$ eV) within the uncertainty. 
On the other hand, it appears that the retrograde BH spin of $a/M=-0.5$ is slightly more favored by observations indicating a relatively higher disk temperature $kT_{\rm in}=34.0$ eV and a slightly lower electron energy $kT_e=61.3$ keV in comparison with the other two cases as seen in Table~3. 
One way to understand this result is the following; in the retrograde case, incoming disk photons are emitted primarily at a larger disk radius according to equation~(15) while the downstream region remains to be formed at small radii (i.e. $r_{\rm sh}/r_g=2.16$). Thus, the emitted photons from the disk are subject to more blueshift in the rest-frame of the downstream plasma. For such Comptonized photons with more blueshift, high electron energy $kT_e$ would not be necessary to counteract against a subsequent gravitational redshift while propagating towards the observer.   
Whereas even lower electron energy $kT_e (\lesssim 61 \rm{keV})$ for $a/M=-0.5$ can be statistically viable as shown in Figure~\ref{fig:f11}b, we find that no plasma accretion in this case is allowed to develop a downstream region that is ``colder" than $kT_e \sim 61$ keV as seen in Figure~\ref{fig:f11}b which is contrasted with the other two cases where the confidence level is unambiguously constrained as shown in Figures~\ref{fig:f9}b and \ref{fig:f10}b.


\begin{deluxetable}{l c ccc}
\tablecaption{Best-fit GRMHD {\tt compsh}  Models$^{\dagger}$ \label{tab:tab3}}
\tabletypesize{\small}
\tablecolumns{10}
\tablewidth{0pt}
\tablehead{
	\colhead{Parameter} & \colhead{Description} &
	\multicolumn{3}{c}{BH Spin $a/M$} \\
	\cline{3-5} \\
	\colhead{} & \colhead{} &
	\colhead{-0.5} & \colhead{0} & \colhead{0.5} 
}
\startdata
$kT_{\rm e}$ [keV]$^{\dagger}$ & Electron Energy & $61.3^{+2.0}_{b}$ & $83.5^{+4.2}_{-2.8}$ & $144.3^{+7.2}_{-6.8}$  \\ 
$kT_{\rm in}$ [eV] & Disk Temperature & $34.0^{+1.3}_{-10.3}$ & $21.7^{+0.64}_{-0.53}$ & $21.6^{+0.46}_{-0.40}$  \\ 
$\theta_{\rm obs}$ [deg] & Inclination Angle & $17.5^{+5.2}_{-1.4}$ & $42.6_{-7.3}^{a}$ & $36.6^{+4.3}_{-3.5}$ \\ 
%
%
$r_{\rm A}^{\rm }/r_g$ & \Alfven Point & $8.27^{a}_{-0.028}$ & $3.28^{+0.002}_{-0.001}$ & $2.48^{+0.001}_{-0.001}$  \\
$r_{\rm F}^{\rm out}/r_g$ & Outer Fast Point & $3.53\pm{0.0001}$ & $3.00\pm{0.0001}$ & $2.35^\pm{0.0001}$  \\
$r_{\rm sh}/r_g$ & Shock Location & $2.16^{+0.004}_{b}$ & $2.69^{+0.002}_{-0.001}$ & $1.99^{+0.005}_{-0.004}$  \\
$r_{\rm F}^{\rm in}/r_g$ & Inner Fast Point & $2.11^{a}_{-0.0003}$ & $2.42^{+0.002}_{-0.003}$ & $1.90^{+0.001}_{-0.001}$  \\
$n_2/n_1$ & Compression Ratio & $1.08^{+0.008}_b$ & $1.10^{+0.002}_{-0.001}$ & $1.82^{+0.01}_{-0.008}$ \\
$\sigma_1$ & Upstream Magnetization & $0.41^{a}_{-0.006}$ & $0.037^{0.001}_{-0.001}$ & $1.07^{+0.06}_{-0.05}$ \\
$\sigma_2$ & Downstream Magnetization & $0.49^{a}_{-0.0005}$ & $0.21^{+0.004}_{-0.002}$ & $2.82^{+0.12}_{-0.11}$ \\
$B_{\phi 2}/B_{\phi 1}$  & Toroidal Field Enhancement & $1.11^{+0.014}_{b}$ & $5.01^{+0.25}_{-0.17}$ & $1.43^{+0.025}_{-0.021}$  \\ \hline
$\chi^2$/dof & & 961.91/944 & 994.77/944 & 1014.64/944 
\enddata \\
\vspace*{0.2cm}
All radii are normalized to the gravitational radius $r_g$. \\
$^{\dagger}$ Assuming that $M_8=1, q=-2, \beta_1=0.01$ and $f_c=1.7$. \\
$^a$ Upper parameter value reached. \\
$^b$ Lower parameter value reached.
\end{deluxetable}

\section{Summary \& Conclusion}

In this work we proposed a novel scenario that a shock-heated downstream flow in GRMHD accretion, injected from near the ISCO, serves as an ideal heating site where accelerated electrons can Compton up-scatter thermal disk photons to imprint the observed soft excess component in the AGN X-ray spectrum. Extending our earlier work on GRMHD standing shock formation, we explored sets of fiducial solutions for accreting plasma in Kerr geometry and studied their physical conditions in terms of liberated energy via shocks. Given a monochromatic blackbody radiation originating from the hottest part of the disk, we calculated Comptonized spectra by making use of the obtained plasma accretion. Our  calculations are all fully relativistic including photon redshift among the disk rest-frame, plasma rest-frame and the distant observer's rest-frame by employing GR ray-tracing approach.      
As a simplistic three-parameter model (i.e. $\theta_{\rm obs}, kT_e$ and $kT_{\rm in}$) besides its normalization, we have constructed a grid of synthetic spectral model for Comptonized component ({\tt compsh}) and  further demonstrated that the model can successfully explain the observed soft excess feature for a typical Seyfert 1 galaxy, Ark~120, as a case study.  

We show that GRMHD plasma in this model begins to plunge in from radii very close to the  ISCO within the standard disk paradigm.
It is found that the downstream region, heated by fast MHD shocks, typically extends out to only a few gravitational radii. While we only consider equatorial accretion for simplicity, plasma in reality is expected to accrete along a closed-loop of poloidal field lines within the ISCO developing shocks as shown in F07 where the vertical height of the shock is systematically found to be $h \lesssim 5 r_g$ for various accretion parameters. Such a very compact and centrally-concentrated region resembles a long-sought physical identity of the putative X-ray ``coronal" region near the BH that is inevitably required for explaining the basis of the major spectral components in AGN X-ray observations \citep[e.g.][]{HM91,Miniutti03,Petrucci04,Nardini11,Fabian15,Keck15,Lohfink15}. For example, \citet{Kara15} has analyzed the broad-band X-ray spectrum of a narrow-line Seyfert galaxy, 1H0707-495, to estimate a very compact coronal source at the height of $h \sim 2 r_g$ above a rapidly-rotating BH in the context of the standard lamp-post model, while its physical identification of the X-ray source in that extreme proximity is yet to be confirmed theoretically.   
Our GRMHD calculations indicate that the formation of such a ``corona" is ubiquitous and almost universal for a wide range of plasma parameter space including BH spin and the $\chi^2$ analysis successfully derived the range of $61.3~{\rm keV} \lesssim kT_e \lesssim 144.3~{\rm keV}$, $21.6~{\rm eV} \lesssim kT_{\rm in} \lesssim 34.0~{\rm eV}$ and  $17.5\degr \lesssim \theta_{\rm obs} \lesssim 42.6\degr$, although the model slightly favors the retrograde BH case ($a/M=-0.5$) as shown in Table~\ref{tab:tab3}.
While a discussion of physical arrangement of such a retrograde AGN system is beyond the scope of this paper, counterrotating plasma (with respect to BH rotation) allows for a larger Comptonizing area due to a larger  ISCO radius \citep[e.g.][]{Chandra83,Bardeen72,Cunningham75}. 
Hence, the resulting composite spectrum from the downstream flow can afford to produce a more diverse spectral shape when integrating over the downstream region. This may be the reason why the  soft excess in data seems to be better accounted for by a retrograde BH case. The current model, however, does not provide a fundamental explanation of such  a retrograde BH spin and it is only empirical in this framework. 
A retrograde BH, allowing for the disk to recede further out, may also be viable with the fact that Ark~120 seems to exhibit   no strong signs for absorption (thus a ``bare" nucleus) which would otherwise be expected to be present in soft X-ray band. We note, however, that the earlier studies of Ark~120 seem to imply a somewhat intermediate BH spin of $a/M \sim 0.5-0.6$ based on the ionized reflection models while with some potential uncertainties \citep[e.g.][]{Garcia14,Matt14}.

In the present work we have only considered moderate BH spin values ($-0.5 \le a/M \le 0.5$) because we found it rather challenging to systematically obtain the solutions for very high BH spin values ($a/M \gtrsim 0.9$) as the size of the downstream region becomes even more compact in radius. A more through parameter search will be a future work.

In comparison with a recent thermal Comptonization scenario proposed by \citet{Petrucci13} for MrK~509, the Comptonizing ``corona" (i.e. $60 \lesssim kT_e \lesssim 150$ keV within the ISCO) in our model is very similar to ``hot corona" (e.g. $kT_{\rm hc} \sim 100$ keV at the innermost region of accretion responsible for the power-law component) but different from ``warm corona" ($kT_{\rm wc} \sim 0.6$ keV exterior to the hot corona responsible for the soft excess) in their model in terms of its geometry and physical characteristics (see their Fig.~10). A critical difference between the two models lies in the fact that the spectral distribution of Comptonizing particles (i.e. electrons) is assumed to be nonthermal (i.e. power-law) as described in \S 3.1. while \citet{Petrucci13} considers thermal distribution. Although our current model focuses only on the production of soft excess through shocks, it is very likely that a local magnetic field activity on the disk surface such as reconnection might play a major role in producing nonthermal continuum in hard X-ray regime.


As has been widely debated to date in the literature, relativistically-blurred reflection model may also be able to explain the observed excess feature \citep[e.g.][]{Ponti10,DeMarco13,Lohfink12,Vasudevan14,Crummy06} while expecting a strong spectral  correlation between the soft and hard X-rays due to their direct coupling in production process.    
In the framework of our current model, on the other hand, the production site of the soft excess component is strictly confined to the innermost plasma accretion set by the shock formation within the ISCO radius and thus   physically disentangled from reflected hard X-rays  expecting no correlation between the two.  In recent studies of broad-band spectroscopies from a number of Seyfert galaxies showing the strong soft excess, however, we note that detailed spectral analyses seem to disfavor  such an expected correlation but indicate a correlation between UV and soft X-ray flux \citep[e.g.][]{Mehdipour11,Petrucci13,Boissay14,Boissay15}. Our model, by definition, is co-aligned with the latter findings. While not definitive yet, our scenario is consistent with those findings at least qualitatively. As discussed by \citet{Matt14}, on the other hand, the actual  correlation could be rudimentary as the soft excess could be present without a pronounced relativistic reflection component, and it takes more effort to draw a conclusion.    

It is found  that Comptonization due to standing shock is very sensitive to the shock property controlled primarily by the downstream energy $kT_e$ while not so significantly dependent on BH spin $a$ in that the calculated spectra are almost generically self-similar regardless of the exact value of BH spin. This may thus be indicative of model degeneracy with BH spin. 
In the present calculations we treat arbitrarily the normalization of Comptonized spectrum as a free parameter. That is, the flux level of the calculated excess component is 
not  determined by the model but is only provided by data. Technically speaking, the  intensity of the excess component must be coupled with plasma property (such as density, temperature and magnetic field strength) and should be  self-consistently calculated  by the model.  It is important, however, to examine how much energy is   physically available in the postshock region in comparison with the observed soft excess flux. To clearly address the modeled excess intensity, a more rigorous consideration of the {\tt compsh} normalization is necessary  as a future work.   While still open to debate how much Comptonized flux can be produced from the small downstream region, one might argue that relativistic beaming (via light bending effect) of seed photons towards small radii could preferentially allow for a sufficient centrally-concentrated illumination \citep[e.g.][]{FK07a}.  This may yield enough Comptonized photons possibly in a scenario analogous to the observed double-peaked H$\alpha$ emission line from AGN accretion disk \citep[e.g.][]{Chen89,Eracleous94,Strateva08}.        

In this work, we assumed a split-monopole field configuration. Although not globally applicable, this is an approximate solution to the trans-field equation (i.e. GR GS-equation) as originally discussed in \citet{BZ77}. 
Interestingly, state-of-the-art numerical simulations in recent years from different groups indicate a characteristic field topology very similar to that described as the split-monopole \citep[][]{Hirose04,Tchekhovskoy09,Contopoulos13}. 
Although a detailed structure and strength of the actual magnetic field at the horizon scale still remains unclear, this approximation is a first step forward to the problem. Nonetheless, the accreting plasma models listed in Table~1 yield the  field strength on the order of $B \sim 10^{3-4}$ G consistent with the known estimates to date \citep[e.g.][]{Krolik99,Wang01,F07}.

%


While treated as fully relativistic under strong gravity, we note that the current model is time-independent based on axisymmetric plasma. This assumption makes it impossible for us to predict any temporal nature of the soft excess considered in this work; e.g. spectral time variabilities associated with shock compression and cooling effects.  
The downstream plasma properties are numerically solved by considering adiabatic (nonradiative) Rankine-Hugoniot jump conditions as a pure mathematical discontinuity with no energy/mass loss. Hence, most of the heat generated at the shock front is advected with the downstream plasma. A more realistic shock process, on the other hand, is most likely accompanied by radiative cooling to some degree in which the postshock plasma temperature may stay comparatively as cool as that of the upstream one as in the isothermal shocks \citep[e.g.][]{Lu98,Das03,F04,FK07b}. 
Radiative dissipation at the shock front could therefore drastically change the subsequent downstream plasma condition which in turn alters the Comptonization process.
In reality, furthermore, accreting plasma may be characterized by a two-temperature gas between electrons and ions \citep[e.g.][]{SLE76,Mahadevan98,Manmoto00} unless the Coulomb coupling between the two is very efficient, whereas in this work we prescribed a single-fluid approximation for simplicity. 
\cite{Becker11} have considered a particle transport process (e.g. bulk advection, spatial diffusion and particle escape) via the effects of the first-order Fermi acceleration across a standing shock. In a more self-consistent scenario such a calculation of diffusive shock acceleration should be incorporated to reflect the energetic outflows/jets from the shock front.
Although all these micro-physics should be addressed and incorporated into more sophisticated calculations by GRMHD simulations for completeness, this is beyond the scope of this paper.   

The other potentially important spectral components associated with magnetic fields include synchrotron radiation, and its Comptonization have also been extensively considered in the literature in the context of black hole binaries such as Cygnus X-1, for example, via nonthermal powerlaw electrons \citep[e.g.][]{WardzinskiZdziarski01,ChakrabartiMandal06}. 
While this is especially important in black hole binaries, 
the characteristic synchrotron frequency in AGNs is estimated to be $\nu_{\rm syn} \simeq 4 \times 10^{10} B_4 \gamma_e^2 ~ \rm{[Hz]} ~ \sim 1.6 \times 10^{-4}$ [eV] where $B_4$ is the field strength in units of $10^4$ G and we take $\gamma_e \sim 1$ in this model. Hence, those Comptonized photons are less likely to be significant to the composite spectrum from the innermost accretion region of AGNs that are considered in this work.


We have analyzed in this work a Seyfert galaxy, Ark~120, to demonstrate that the proposed model successfully describes the observed soft excess feature within the framework of a simplistic accretion model. As there is a number of archival X-ray data available mainly from typical narrow-line Seyfert AGNs and PG quasars that also exhibit strong excess components \citep[e.g.][for 20 to 30 AGNs]{Crummy06} in the {\it XMM-Newton}/EPIC-pn observations, we will extend the current study to those available for a more systematic analysis. 
Despite a simplistic prescription of the proposed scenario based on GRMHD shock formation, 
our Comptonization model is successful in describing the observed soft excess feature in Ark~120. We thus find the current result to be an encouraging first step towards the next level where additional relevant physics are employed to make the model more physically self-consistent and promising.

As a next-generation X-ray observatory, we anticipate advanced new X-ray missions, such as {\it Athena}, 
to contribute
significantly to this study particularly with the high-resolution X-ray microcalorimeter spectrometer  by providing more detailed spectra simultaneously on the soft and hard X-ray 
components. The expected data will thus help  differentiate various (fundamentally) distinct models presented today 
and further clarify the current ambiguous views concerning the observed soft excess in the immediate circumnuclear region of AGNs.

\acknowledgments
The authors acknowledge the anonymous referee for useful comments and questions. KF is grateful to Omer Blaes for his insightful suggestions about the disk simulations and Rozenn Boissay for a number of useful comments. A part of this work was conducted while at KITP of UCSB and also supported in part by the 4-VA Collaborative at James Madison University.


\begin{thebibliography}{999}

%
%
%
%
%
%



\bibitem[Acharya et al.(2002)]{Acharya02} Acharya, K., Chakrabarti, S. K. \& Molteni, D. 2002, JApA, 23, 155
\bibitem[Armitage et al.(2001)]{Armitage01} Armitage, P. J., Reynolds, C. S., \& Chiang, J. 2001, ApJ, 548, 868
\bibitem[Arnaud(1996)]{Arnaud96} Arnaud, K. A. 1996, in ASP Conf. Ser. 101, Astronomical Data Analysis Software and Systems V, ed. G. H. Jacoby \& J. Barnes (San Francisco,
CA: ASP), 17
\bibitem[Bardeen et al.(1972)]{Bardeen72} Bardeen, J. M., Press, W. H., \& Teukolsky, S. A. 1972, \apj, 178, 347
\bibitem[Baring(1997)]{Baring97} Baring, M. G. 1997, in Very High Energy Phenomena in the Universe, ed.
Y. Giraud-Heraud \& J. Tran Thanh Vanastro ( Paris: Editions Fronti\`{e}res),
97
\bibitem[Becker et al.(2011)]{Becker11} Becker, P. A., Das, S. \& Le, T. 2011, \apj, 743, 47
\bibitem[Boissay et al.(2014)]{Boissay14} Boissay, R., Paltani, S., Ponti, G., Bianchi, S., Cappi, M., Kaastra, J. S., Petrucci, P.-O., Arav, N., Branduardi-Raymont, G., Costantini, E., Ebrero, J., Kriss, G. A., Mehdipour, M., Pinto, C. \& Steenbrugge, K. C. 2014, \aap, 567, 44 
\bibitem[Boissay et al.(2015)]{Boissay15} Boissay, R., Paltani, S. \& Ricci, C. 2015, The Extremes of Black Hole Accretion, Proceedings of the conference 
\bibitem[Boller et al.(1996)]{Boller96} Boller, T., Brandt, W. N., \& Fink, H. 1996, \aap, 305, 53 
\bibitem[Blandford \& Znajek(1977)]{BZ77} Blandford, R. D., \& Znajek, R. L. 1977, MNRAS, 179, 433
\bibitem[Brandt et al.(1993)]{Brandt93} Brandt, W. N., Fabian, A. C., Nandra, K., \&  Tsuruta, S. 1993, \mnras, 265, 996
\bibitem[Chakrabarti(1990)]{C90} Chakrabarti, S. K. 1990, Theory of Transonic Astrophysical Flows (Singapore: World Scientific)
\bibitem[Chakrabarti(1995)]{C95} Chakrabarti, S. K. 1995, Seventeeth Texas Symposium on Relativistic Astrophysics and Cosmology, Annuals of the New York Academy of Sciences, Vol. 759 (New York, NY: The New York Academy of Sciences), p.546
\bibitem[Chakrabarti \& Titarchuk(1995)]{CT95} Chakrabarti, S. \& Titarchuk, L. G. 1995, ApJ, 455, 623 
\bibitem[Chakrabarti \& Mandal(2006)]{ChakrabartiMandal06} Chakrabarti, S. K. \& Mandal, S. 2006, \apj, 642, L49
\bibitem[Chandrasekhar(1983)]{Chandra83} Chandrasekhar, S. 1983, The Mathematical Theory of Black Holes (Oxford:
Oxford Univ. Press)
\bibitem[Chen et al.(1989)]{Chen89} Chen, K., Halpern, J. P. \& Filippenko, A. V. 1989, \apj, 339, 742
\bibitem[Contopoulos, Kazanas, \& Papadopoulos(2013)]{Contopoulos13} Contopoulos, I., Kazanas, D., \& Papadopoulos, D. B. 2013, \apj, 765, 113
\bibitem[Crummy et al.(2006)]{Crummy06} 	
Crummy J., Fabian A.C., Gallo L., \& Ross R.R. 2006, MNRAS, 365, 1067
\bibitem[Cunningham(1975)]{Cunningham75} Cunningham, C. T. 1975, \apj, 202, 788
\bibitem[Das et al.(2003)]{Das03} Das, T. K., Pendharkar, J. K. \& Mitra, S. 2003, \apj, 592, 1078
\bibitem[Das \& Chakrabarti(2007)]{Das07} Das, S. \& Chakrabarti, S. K. 2007, MNRAS, 374, 729 
\bibitem[Davis et al.(2005)]{Davis05} Davis, S. W., Blaes, O. M., Hubeny, I., \& Turner, N. J. 2005, \apj, 621, 372
\bibitem[De Marco et al.(2013)]{DeMarco13} 
De Marco, B., Ponti, G., Cappi, M., Dadina, M., Uttley, P., Cackett, E. M., Fabian, A. C., \&  Miniutti, G. 2013, \mnras, 431, 2441
\bibitem[Dickey \& Lockman(1990)]{Dickey90} Dickey, J. M. \& Lockman, F. J. 1990, ARA\&A, 28, 215
\bibitem[Di Gesu et al.(2014)]{DiGesu14} Di Gesu, L., CostanMni, E., Piconcelli, E., et al. 2014, A\&A, 563, A95
\bibitem[Done et al.(2012)]{Done12} 
Done, C., Davis, S. W., Jin, C., Blaes, O., \& Ward, M. 2012, \mnras, 420, 1848
\bibitem[Droege \& Schlickeiser(1986)]{DroegeSchlickeiser86} 
Droege, W. \& Schlickeiser, R. 1986, \apj, 305, 909
\bibitem[Eracleous \& Halpern(1994)]{Eracleous94} Eracleous, M. \& Halpern, J. P. 1994, \apjs, 90, 1
\bibitem[Fabian et al.(1989)]{Fabian89} Fabian, A. C., Rees, M. J., Stella, L., \&  White, N. E. 1989, \mnras, 238, 729
\bibitem[Fabian et al.(2004)]{Fabian04} Fabian, A. C., Miniutti, G., Gallo, L., Boller, Th., Tanaka, Y., Vaughan, S. \& Ross, R. R. 2004, \mnras, 353, 1071
\bibitem[Fabian et al.(2009)]{Fabian09} Fabian, A. C., Zoghbi, A., Ross, R. R., et al. 2009, Nature, 459, 540
\bibitem[Fabian et al.(2015)]{Fabian15} Fabian, A. C., Lohfink, A., Kara, E., Parker, M. L., Vasudevan, R., \& Reynolds, C. S. 2015, \mnras, 451, 4375
\bibitem[Fermi(1949)]{Fermi49} Fermi, E. 1949, Phys. Rev., 75, 1169



\bibitem[Fragile et al.(2007)]{Fragile07} Fragile, P. C., Blaes, O. M., Anninos, P., \& Salmonson, J. D. 2007, ApJ, 668, 417
\bibitem[Fragile \& Blaes(2008)]{FB08} Fragile, P. C. \& Blaes, O. M. 2008, ApJ, 687, 757 

\bibitem[Frank et al.(1992)]{Frank92} Frank, J., King, A., \& Raine, D. 1992, Accretion Power in Astrophysics (2nd ed.; Cambridge: Cambridge Univ. Press)
\bibitem[Fukumura et al.(2004)]{F04} Fukumura, K. \& Tsuruta, S. 2004, \apj, 611, 964
\bibitem[Fukumura et al.(2007)]{F07} Fukumura, K., Takahashi, M. \& Tsuruta, S. 2007, \apj, 657, 415 (F07)
\bibitem[Fukumura \& Kazanas(2007a)]{FK07a} Fukumura, K. \& Kazanas, D. 2007a, \apj, 664, 14
\bibitem[Fukumura \& Kazanas(2007b)]{FK07b} Fukumura, K. \& Kazanas, D. 2007b, \apj, 669, 85

\bibitem[Garc\'{i}a et al.(2014)]{Garcia14} Garc\'{i}a, J., Dauser, T., Lohfink, A., Kallman, T. R., Steiner, J. F., McClintock, J. E., Brenneman, L., Wilms, J., Eikmann, W., Reynolds, C. S. \& Tombesi, F. 2014, \apj, 782, 76

\bibitem[Generozov et al.(2014)]{Generozov14} Generozov, A., Blaes, O., Fragile, P. C., \& Henisey, K. B. 2014, \apj, 780, 81
\bibitem[Gierli\'{n}ski \& Done(2004)]{GierlinskiDone04} Gierli\"{n}ski, M. \& Done, C. 2004, \mnras, 349, 7
\bibitem[Gieseler \& Jones(2000)]{GieselerJones00} Gieseler, U. D. J., \& Jones, T. W. 2000, A\&A, 357, 1133
\bibitem[Goodrich(1989)]{Goodrich89} Goodrich, R. W. 1989, ApJ, 342, 224
\bibitem[Haardt \& Maraschi(1991)]{HM91} Haardt, F., \& Maraschi, L. 1991, ApJ, 380, L51
\bibitem[Hirose et al.(2004)]{Hirose04} Hirose, S., Krolik, J. H., De Villiers, J.-P. \&  Hawley, J. F. 2004, ApJ, 606, 1083
\bibitem[Hollywood \& Melia(1997)]{Hollywood97} 
Hollywood, J. M. \& Melia, F. 1997, \apjs, 112, 324
\bibitem[Kara et al.(2015)]{Kara15} 
Kara, E., Fabian, A. C., et al. 2015, \mnras, 449, 234
\bibitem[Kato et al.(2008)]{Kato08} Kato, S., Fukue, J., \& Mineshige, S. 2008, Black Hole Accretion Disks (2nd ed.; Kyoto: Kyoto Univ. Press) 
\bibitem[Keck et al.(2015)]{Keck15} Keck, M. L. et al. 2015, \apj, 806, 149
\bibitem[Koide et al.(1998)]{Koide98} Koide, Shinji; Shibata, Kazunari; Kudoh, Takahiro, 1998, \apj, 495, 63
\bibitem[Koide et al.(2000)]{Koide00} Koide, Shinji; Meier, David L.; Shibata, Kazunari; Kudoh, Takahiro, 2000, \apj, 536, 668
\bibitem[Krolik(1999)]{Krolik99} Krolik, J. H. 1999, Active Galactic Nuclei (Princeton: Princeton Univ. Press)
\bibitem[Leighly(1999a)]{Leighly99a} Leighly, K. M. 1999a, \apjs, 125, 297
\bibitem[Leighly(1999b)]{Leighly99b} Leighly, K. M. 1999b, \apjs, 125, 317
\bibitem[Kojima(1991)]{Kojima91} Kojima, Y. 1991, \mnras, 250, 629
\bibitem[Laor et al.(1997)]{Laor97} Laor, A., Fiore, F., Elvis, M., Wilkes, B. J., \& McDowell, J. C. 1997, \apj, 477, 93
\bibitem[Le \& Becker(2005)]{LeBecker05} Le, T. \& Becker, P. A. 2005, \apj, 632, 476
\bibitem[Li et al.(2005)]{Li05} 
Li, L.-.X., Zimmerman, E. R., Narayan, R., \& McClintock, J. E. 2005, \apjs, 157, 335
\bibitem[Lohfink et al.(2012)]{Lohfink12} 
Lohfink, A. M., Reynolds, C. S., Miller, J. M., Brenneman, L. W., Mushotzky, R. F., Nowak, M. A. \& Fabian, A. C. 2012, \apj, 758, 67
\bibitem[Lohfink et al.(2015)]{Lohfink15} 
Lohfink, A. M., et al. 2015, \apj, 814, 24
\bibitem[Lu et al.(1997)]{Lu97} Lu, J.-F., Yu, K. N., Yuan, F., \& Young, E. C. M. 1997, \aap, 321, 665
\bibitem[Lu \& Yuan(1998)]{Lu98} Lu, J.-F. \& Yuan, F. 1998, \mnras, 295, 66
\bibitem[Mahadevan(1998)]{Mahadevan98} Mahadevan, R. 1998, Nature, 394, 651
\bibitem[Manmoto(2000)]{Manmoto00} Manmoto, T. 2000, \apj, 534, 734
\bibitem[Matt et al.(2014)]{Matt14} Matt, G., et al. 2014, \mnras, 439, 3016
\bibitem[Mehdipour et al.(2011)]{Mehdipour11} Mehdipour, M., Branduardi-Raymont, G., Kaastra, J. S., Petrucci, P. O., Kriss, G. A., Ponti, G., Blustin, A. J., Paltani, S., Cappi, M., Detmers, R. G., \& Steenbrugge, K. C. 2011, \aap, 534, 39
\bibitem[Michel(1973)]{Michel73} Michel, F. C. 1973, ApJ, 180, L133
\bibitem[Middleton et al.(2007)]{Middleton07} Middleton, M., Done, C. \& Gierli\"{n}ski, M. 2007, \mnras, 381, 1426
\bibitem[Mineshige et al.(2000)]{Mineshige00} Mineshige, S., Kawaguchi, T., Takeuchi, M. \& Hayashida, K. 2000, \pasj, 52, 499
\bibitem[Miniutti et al.(2003)]{Miniutti03} Miniutti, G., Fabian, A. C., Goyder, R., \& Lasenby, A. N. 2003, MNRAS, 344, L22
\bibitem[Miniutti \& Fabian(2004)]{Miniutti04} Miniutti, G. \& Fabian, A. C. 2004, \mnras, 349, 1435 
\bibitem[Mitsuda et al.(1984)]{Mitsuda84} 
Mitsuda, K., Inoue, H., Koyama, K., Makishima, K., Matsuoka, M., Ogawara, Y., Suzuki, K., Tanaka, Y., Shibazaki, N., \& Hirano, T. 1984, \pasj, 36, 741

\bibitem[Molteni et al.(1996)]{Molteni96} Molteni, D., Sponholz, H., Chakrabarti, S. K. 1996, ApJ, 457, 805
\bibitem[Molteni et al.(1999)]{Molteni99} Molteni, D., Gerardi, G., Valenza, M. A. \&  Lanzafame, G. 1999, Observational Evidence for the Black Holes in the Universe, Conference held in Calcutta, January 11-17th, p.83

\bibitem[Morales et al.(2014)]{Morales14} Morales T. D., Fragile, P. C., Zhuravlev, V. V., \& Ivanov, P. B. 2014, ApJ, 796, 103

\bibitem[Nagakura \& Yamada(2008)]{NagakuraYamada08} Nagakura, H. \& Yamada, S. 2008, \apj, 689, 391 
\bibitem[Nardini et al.(2011)]{Nardini11} Nardini, E., Fabian, A. C., Reis, R. C., \& Walton, D. J. 2011, \mnras, 410, 1251
\bibitem[Nobuta \& Hanawa(1994)]{NobutaHanawa94} Nobuta, K. \& Hanawa, T. 1994, \pasj, 46, 257
\bibitem[Noda et al.(2013)]{Noda13} 	
Noda, H., Makishima, K., Nakazawa, K., Uchiyama, H., Yamada, S., \& Sakurai, S. 2013, \pasj, 65, 4
\bibitem[Okuda et al.(2004)]{Okuda04} Okuda, T., Teresi, V., Toscano, E., \& Molteni, D. 2004, PASJ, 56, 547 	
\bibitem[Okuda et al.(2007)]{Okuda07} Okuda, Y., Teresi, V. \& Molteni, D. 2007, MNRAS, 377, 1431
\bibitem[Osterbrock \& Pogge(1985)]{OsterbrockPogge85} Osterbrock, D. E. \& Pogge, R. W. 1985, ApJ, 297, 166	

\bibitem[Peterson et al.(2004)]{Peterson04} 
Peterson, B. M., Ferrarese, L., Gilbert, K. M., Kaspi, S., Malkan, M. A., Maoz, D., Merritt, D., Netzer, H., Onken, C. A., Pogge, R. W., Vestergaard, M., \& Wandel, A. 2004, \apj, 613, 682
\bibitem[Petrucci et al.(2004)]{Petrucci04} Petrucci, P. O., Maraschi, L., Haardt, F. \&  Nandra, K. 2004, \aap, 413, 477
\bibitem[Petrucci et al.(2013)]{Petrucci13} Petrucci, P.-O., Paltani, S., Malzac, J., et al. 2013, A\&A, 549, A73
\bibitem[Ponti et al.(2010)]{Ponti10} 	
Ponti, G., Gallo, L. C., Fabian, A. C., Miniutti, G., Zoghbi, A., Uttley, P., Ross, R. R., Vasudevan, R. V., Tanaka, Y. \& Brandt, W. N. 2010, \mnras, 406, 2591
\bibitem[Pu et al.(2015)]{Pu15} Pu, H.-Y., Nakamura, M., Hirotani, K., Mizuno, Y., Wu, K. \&  Asada, K. 2015, \apj, 801, 56

\bibitem[Ross et al.(1992)]{Ross92} Ross, R. R., Fabian, A. C. \& Mineshige, S. 1992, \mnras, 258, 189
\bibitem[Ross \& Fabian(2005)]{RossFabian05} Ross R.R. \& Fabian A.C. 2005, MNRAS, 358, 211
\bibitem[Schurch \& Done(2006)]{SD06} Schurch, N. J. \& Done, C. 2006, \mnras, 371, 81
\bibitem[Schurch \& Done(2008)]{SD08} Schurch, N. J. \& Done, C. 2008, \mnras, 386, 1
\bibitem[Shakura \& Sunyaev(1973)]{SS73} Shakura, N. I.; Sunyaev, R. A. 1973, \aap, 24, 337


\bibitem[Shapiro et al.(1976)]{SLE76} Shapiro, S. L., Lightman, A. P., \& Eardley, D. M. 1976, ApJ, 204, 187

\bibitem[Shimura \& Takahara(1995)]{ShimuraTakahara95} Shimura, T. \& Takahara, F. 1995, \apj, 445, 780
\bibitem[Stark et al.(1992)]{Stark92} Stark, Antony A., Gammie, C. F., Wilson, R. W., Bally, J., Linke, R. A., Heiles, C., \& Hurwitz, M. 1992, \apjs, 79, 77
\bibitem[Strateva et al.(2008)]{Strateva08} Strateva, I. V., Brandt, W. N., Eracleous, M. \& Garmire, G. 2008, \apj, 687, 869
\bibitem[Takahashi et al.(2002)]{T02} Takahashi, M., Rilett, D., Fukumura, K. \& Tsuruta, S. 2002, \apj, 572, 950 (T02)
\bibitem[Takahashi et al.(2006)]{T06} Takahashi, M., Goto, J., Fukumura, K., Rilett, D. \& Tsuruta, S. 2006, \apj, 645, 1408 (T06) 
\bibitem[Takahashi \& Takahashi(2010)]{T10} Takahashi, M. \& Takahashi, R. 2010, \apj, 714, 176
\bibitem[Tchekhovskoy, McKinney, \& Narayan(2009)]{Tchekhovskoy09} Tchekhovskoy, A., McKinney, J. C. \& Narayan, R. 2009, \apj, 699, 1789 
\bibitem[Tchekhovskoy et al.(2011)]{TNM11} Tchekhovskoy, A., Narayan, R. \& McKinney, J. C. 2011, MNRAS, 418, 79 
\bibitem[Titarchuk et al.(1996)]{TMK96} Titarchuk, L., Mastichiadis, A. \& Kylafis, N. D. 1996, A\&AS, 120, 171
\bibitem[Turner \& Pounds(1989)]{TurnerPounds89} Turner, T. J. \& Pounds, K. A. 1989, \mnras, 240, 833
\bibitem[Vasudevan et al.(2014)]{Vasudevan14} Vasudevan, R. V., Mushotzky, R. F., Reynolds, C. S., Fabian, A. C., Lohfink, A. M., Zoghbi, A., Gallo, L. C. \& Walton, D. 2014, \apj, 785, 30
\bibitem[Vaughan et al.(2004)]{Vaughan04} Vaughan, S., Fabian, A. C., Ballantyne, D. R., De Rosa, A., Piro, L., \& Matt, G. 2004, \mnras, 351, 193
\bibitem[Wald(1974)]{Wald74} Wald, R. M. 1974, Phys. Rev. D, 10, 1680
\bibitem[Walter \& Fink(1993)]{WalterFink93} Walter, R. \& Fink, H. H. 1993, \aap, 274, 105
\bibitem[Wandel et al.(1999)]{WandelPetersonMalkan99} Wandel, A.; Peterson, B. M.; Malkan, M. A. 1999, \apj, 526, 579
\bibitem[Wang et al.(2001)]{Wang01} Wang, T. G., Matsuoka, M., Kubo, H., Mihara, T., \& Negoro, H. 2001, ApJ,
554, 233
\bibitem[Wardzi\'{n}ski \& Zdziarski(2001)]{WardzinskiZdziarski01} Wardzi\'{n}ski, G. \& Zdziarski, A. A. 2001, MNRAS, 325, 963
\bibitem[Wilkins et al.(2015)]{Wilkins15} Wilkins, D. R., Gallo, L. C., Grupe, D., Bonson, K., Komossa, S. \& Fabian, A. C. 2015, MNRAS, 454, 4440 
\bibitem[Zhong \& Wang(2013)]{ZW13}  Zhong, X. \& Wang, J. 2013, \apj, 773, 23


%


\end{thebibliography}
\end{document}